\UseRawInputEncoding

\documentclass[nofootinbib,prx,aps,twocolumn,floatfix,10pt,longbibliography,superscriptaddress,notitlepage]{revtex4-1}
\usepackage[english]{babel}
\usepackage{breakcites}
\usepackage{listings}

\usepackage[T1]{fontenc}
\usepackage{textcomp}
\usepackage{gensymb,braket}

\usepackage{graphicx}
\usepackage{braket}
\usepackage{color}
\usepackage{epstopdf}
\usepackage{mathtools}
\usepackage{amsmath}
\usepackage{amssymb}
\usepackage{float}
\usepackage{comment}
\usepackage{amsmath}
\usepackage{multirow}

\usepackage{amsfonts}
\usepackage{bm}
\usepackage{bbold}

\usepackage{color}

\graphicspath{{./figs/}}
\usepackage{gensymb}
\usepackage[colorinlistoftodos,prependcaption]{todonotes}
\usepackage{xargs}  
\newcommandx{\unsure}[2][1=]{\todo[linecolor=red,backgroundcolor=red!25,bordercolor=red,#1]{#2}}
\newcommandx{\change}[2][1=]{\todo[linecolor=blue,backgroundcolor=blue!25,bordercolor=blue,#1]{#2}}
\newcommandx{\info}[2][1=]{\todo[linecolor=OliveGreen,backgroundcolor=OliveGreen!25,bordercolor=OliveGreen,#1]{#2}}
\newcommandx{\improvement}[2][1=]{\todo[linecolor=Plum,backgroundcolor=Plum!25,bordercolor=Plum,#1]{#2}}
\newcommandx{\thiswillnotshow}[2][1=]{\todo[disable,#1]{#2}}
\newcommandx{\greencom}[2][1=]
{\todo[inline, color=green!40,#1]{#2}}
\newcommandx{\bluecom}[2][1=]
{\todo[inline, color=blue!40,#1]{#2}}

\usepackage[colorlinks]{hyperref}
\definecolor{winered}{rgb}{0.5,0,0}
\hypersetup
{
colorlinks=true,
linkcolor=winered,
urlcolor={winered},
filecolor={winered},
citecolor={winered},
allcolors={winered}
}

\usepackage{letltxmacro}
\LetLtxMacro{\ORIGselectlanguage}{\selectlanguage}
\makeatletter
\DeclareRobustCommand{\selectlanguage}[1]{%
  \@ifundefined{alias@\string#1}
    {\ORIGselectlanguage{#1}}
    {\begingroup\edef\x{\endgroup
       \noexpand\ORIGselectlanguage{\@nameuse{alias@#1}}}\x}%
}
\newcommand{\definelanguagealias}[2]{%
  \@namedef{alias@#1}{#2}%
}
\makeatother

\definelanguagealias{en}{english}
\definelanguagealias{EN}{english}

\definecolor{myred}{rgb}{0.6350, 0.0780, 0.1840}
\definecolor{myblue}{rgb}{0.1, 0.1, 0.7}
\definecolor{mygreen}{rgb}{0.1, 0.45, 0.1}

\graphicspath{{figure/}}

\begin{document}

%\title{
%Quasinormal modes of index-modulated microring  resonators close to an exceptional point and chiral power flow from linearly polarized dipole %emitters
%or
\title{
Quasinormal mode analysis of chiral power flow from linearly polarized dipole emitters
coupled to index-modulated microring  resonators close to an exceptional point
} 
\author{Juanjuan Ren}
\email{jr180@queensu.ca}
\affiliation{Department of Physics, Engineering Physics, and Astronomy, Queen's
University, Kingston, Ontario K7L 3N6, Canada}
%         %
\author{Sebastian Franke} 
\email{sebastian.r.franke@gmail.com}
\affiliation{Technische Universit\"at Berlin, Institut f\"ur Theoretische Physik,
Nichtlineare Optik und Quantenelektronik, Hardenbergstra{\ss}e 36, 10623 Berlin, Germany}
\affiliation{Department of Physics, Engineering Physics, and Astronomy, Queen's University, Kingston, Ontario K7L 3N6, Canada}
  \author{Stephen Hughes}
  \email{shughes@queensu.ca}
 \affiliation{Department of Physics, Engineering Physics, and Astronomy, Queen's University, Kingston, Ontario K7L 3N6, Canada}

\begin{abstract}

Chiral emission can be achieved from a circularly polarized dipole emitter in a nanophotonic structure that possess special polarization properties such as a polarization singularity, namely with right or left circularly polarization (C-points). Recently, 
%Ref.~\onlinecite{chen_revealing_2020}
 Chen {\it et.~al.} [Nature Physics 16, 571 (2020)] demonstrated the surprising result of  chiral radiation from a linearly-polarized (LP) dipole emitter, and argued that this effect is caused by a decoupling with the underlying eigenmodes of a non-Hermitian system, working at an exceptional point (EP).
Here we present a quasinormal mode (QNM) approach to model a similar index-modulated ring resonator working near an EP and show the same unusual chiral power flow properties from LP emitters, in direct agreement with the experimental results.
We explain these results quantitatively  without invoking the interpretation of a missing dimension (the Jordan vector) and a decoupling from the cavity eigenmodes, since the correct eigenmodes are the QNMs which explain the chiral emission using only two cavity modes.
By coupling a LP emitter with the
dominant two QNMs of the ring resonator, we show how the chiral emission depend on the position and orientation of the emitter, which is also verified by the excellent agreement with respect to the power flow between the QNM theory and full numerical dipole solutions. 
We also show how a normal mode solution will fail to capture the correct chirality since it 
does not take into account the essential QNM phase. 
Moreover, we  demonstrate how one can achieve frequency-dependent chiral emission, and replace
lossy materials with gain materials in the index modulation to reverse the chirality.
\end{abstract}
\maketitle 

\section{Introduction}
Controlling the photon emission of a quantum dipole emitter is
an important requirement for enabling on-demand quantum light sources, which can be achieved by tailoring the surrounding electromagnetic environment (local density of states).
General photonic structures to achieve this local density of states (LDOS) control~\cite{vahala_optical_2003} include plasmonic resonators/films~\cite{maier_plasmonics:_2007,chang_quantum_2006,benson_assembly_2011,jacob_plasmonics_2011,PhysRevB.85.075303,tame_quantum_2013,lian_efficient_2015,Chikkaraddy2016,ren_evanescent-vacuum-enhanced_2017}, dielectric/metallic waveguides~\cite{tong_subwavelength-diameter_2003,tong_single-mode_2004,akimov_generation_2007,yalla_efficient_2012,yalla_cavity_2014,kato_strong_2015}, and photonic crystal waveguides/cavities~\cite{akahane_high-q_2003,yoshie_vacuum_2004,yalla_cavity_2014,goban_superradiance_2015} with strongly localized fields or/and high quality factors. 
In recent decades, unidirectional emission for circular polarized emitters placed close to chiral points in nanophotonic waveguides have been under intense investigation~\cite{Sllner2015,le_feber_nanophotonic_2015,lodahl_chiral_2017,zhang_chiral_2019}, finding various applications~\cite{young_polarization_2015}, such as isolators~\cite{jalas_what_2013,sayrin_nanophotonic_2015} and non-reciprocal networks~\cite{kimble_quantum_2008}. 
This unidirectional dipole coupling is possible when one couples
a spin-polarized dipole emitter
at a so-called C-point~\cite{young_polarization_2015}, which is a polarization point that exhibits right or left circular polarization.

Certain classes of resonators can also support the coupling between circulating modes that propagate with a specific chirality.
For example, optical whispering-gallery-type microcavities, such as microspheres, microcylinders, microdisks, microrings, microtoroids, microtubes and microbubbles, that support whispering gallery modes (WGMs), which naturally appear in the form of a pair of WGMs (degenerate modes) that propagate with different directions, one clockwise (CW) and the other counter clockwise (CCW) in the azimuthal direction~\cite{leung_completeness_1994,mazzei_controlled_2007,teraoka_resonance_2009,righini_whispering_2011,schunk_identifying_2014,cognee_cooperative_2019}. 
Similar to the aforementioned waveguides, embedding a circular polarized emitter to such resonators (at certain points) would result in chiral emission, and high Purcell factors up to several hundreds 
have been predicted~\cite{martin-cano_chiral_2019}.

With the rapid development of nanophotonics fabrication and the understanding of parity-time (PT) symmetry theory, the introduction of  gain materials to WGM resonators allows one to explore the 
%corresponding 
optical PT symmetry properties~\cite{peng_paritytime-symmetric_2014,chang_paritytime_2014,chen_parity-time-symmetric_2018}, including exceptional points (EPs), where two or more modes (for both eigenvalues and eigenfunctions) coalesce into a single mode~\cite{berry_physics_2004,heiss_exceptional_2004,heiss_physics_2012,PhysRevX.6.021007,miri_exceptional_2019,chen_sensitivity_2019,jin_hybrid_2020,chen_revealing_2020}.
Typical designs consist of two optical resonators, where one is made of lossy material and the other one is an amplifying material~\cite{peng_paritytime-symmetric_2014,chang_paritytime_2014,chen_exceptional_2017,chen_parity-time-symmetric_2018,ren_quasinormal_2021,franke_quantized_2021}, which has potential applications for unidirectional propagation~\cite{peng_paritytime-symmetric_2014,chang_paritytime_2014}, high sensitive sensing~\cite{chen_parity-time-symmetric_2018,wiersig_review_2020}, and high performance lasing~\cite{peng_loss-induced_2014,hodaei_parity-time-symmetric_2014,hayenga_electrically_2020}.
Specially, for a simple single WGM resonator, one could already couple  two degenerate WGMs together and even  select the desired mode, by breaking the symmetry, e.g., by introducing a scattering nanoparticle~\cite{mazzei_controlled_2007,peng_chiral_2016,wiersig_sensors_2016,cognee_cooperative_2019}, or modulating the refractive index (with gain or without gain) along the azimuthal direction~\cite{cai_integrated_2012,feng_single-mode_2014,miao_orbital_2016,chen_revealing_2020}.

Very recently, a striking experimental observation was reported by
Chen {\it et. al.}~\cite{chen_revealing_2020}, where  
unidirectional radiation from a linearly polarized (LP) emitter placed at a special position in a non-Hermitian WGM resonator system
was reported, namely chiral emission without the need for a circular polarized emitter. The authors argued that the chiral radiation field (CW direction) of an emitter could become fully decoupled from the underlying eigenmode (CCW) of the ring resonator, which is different from the traditional wisdom that the emitter will interact with the eigenstate(s) of the embedding environment. 
They interpreted this observation as
being 
caused by an emitter that is coupled to the ``missing dimension'' (Jordan vector) when the system is working at an EP.

In this work, we  explain this
effect and the  mysterious chiral power flow, by employing
a  quasinormal mode (QNM) theory
 of a refractive index modulated microring resonator (similar to the ones shown in Ref.~\onlinecite{chen_revealing_2020,martin-cano_chiral_2019}), where the QNMs are the
natural open cavity modes~\cite{lai_time-independent_1990,leung_completeness_1994,leung_time-independent_1994,leung_completeness_1996,lee_dyadic_1999,kristensen_generalized_2012,sauvan_theory_2013,kristensen_modes_2014,ge_design_2014,PhysRevA.98.043806,PhysRevX.7.021035,lalanne_light_2018,kristensen_modeling_2020}. We explain the chiral power flow from a  LP emitter, 
solely in terms of the coupling to the main two QNMs,
and our results are fully consistent with  related experiments~\cite{chen_revealing_2020}.
However, our interpretation is drastically different to the one reported in Ref.~\onlinecite{chen_revealing_2020}. Instead of a decoupling with the eigenmodes and coupling with the missing dimension (Jordan vector), we find that the unusual chiral power flow can be quantitatively explained by a coupling of the LP emitters with the main two  QNMs of index-modulated microring resonators.
%, even in the case %($\delta n_{\rm Re}=\delta n_{\rm Im}$, the refractive index changes in the real and imaginary parts are equal at the required regions)
%claimed as EP condition in %Ref.~\onlinecite{chen_revealing_2020}.

Moreover, we find that this chiral radiation emission can be dependent on the position, orientation (polarization) and the frequency of the introducing LP emitter, which is verified by the excellent agreement between our QNM solutions and full numerical dipole solutions, i.e., the solution to the classical Maxwell equations with a dipole source.
In contrast, using a more usual
normal mode (NM) solutions will fail to predict the correct power flow, since it neglects the importance of the QNM phase.
In addition, adding material gain instead of loss will reverse the chirality, though the spontaneous emission (SE) rate there need to be fixed through a quantum mechanical correction term~\cite{franke_fermis_2021,ren_quasinormal_2021}.

The rest of our paper is organized as follows:
In Sec.~\ref{sec: theory}, 
we describe the main theory, covering QNMs, Green functions, QNM  expansion for the Green function and an analytical expression of Poynting vectors from QNMs in Sec.~\ref{sec: QNMs}. In Sec.~\ref{sec: NMs}, we also show the more usual  NM expansion. To check the validity of the QNMs/NMs solutions, the full numerical dipole method
%\blue{SF: Sometimes, we use 'full dipole' and sometimes we use 'full numerical dipole', this should be made consistent!} 
is introduced in Sec.~\ref{sec: full} for a rigorous numerical check using the full Maxwell equations. Sec.~\ref{sec: resultsQNMs} shows the two QNMs found for the resonators of interest, including the QNM distributions and complex eigenfrequencies. The generalized Purcell factors from these QNMs are then shown in Sec.~\ref{sec: Purcellfactors}, which show excellent agreement with full numerical dipole results.
As the main feature of the working resonators, the position and orientation (polarization) dependent chiral radiation is discussed in Sec.~\ref{sec: chiralflow}, where we find that the QNM solutions show quantitative agreement with full numerical dipole results, while the NM solutions completely fail. 
More general discussions are included in Sec.~\ref{sec: discussion}, including frequency dependent chirality, chirality with gain materials, and chirality of ring resonator without modulation. 
Finally, in Sec.~\ref{sec: conclusions}, we give a summary and our conclusions.

\section{Theory}\label{sec: theory}

\subsection{Quasinormal mode theory, Green functions and analytical power flow from a dipole}\label{sec: QNMs}
In recent decades, QNM theory~\cite{lai_time-independent_1990,leung_completeness_1994,leung_time-independent_1994,leung_completeness_1996,lee_dyadic_1999,kristensen_generalized_2012,sauvan_theory_2013,kristensen_modes_2014,ge_design_2014,PhysRevA.98.043806,PhysRevX.7.021035,lalanne_light_2018,kristensen_modeling_2020} has shown great success for describing the interaction between dipole emitters and open systems. The QNM eigenfunctions $\tilde{\mathbf{f}}_{\mu}$ are obtained by solving the (source-free) Helmholtz equation 
\begin{equation}\label{smallf}
\boldsymbol{\nabla}\times\boldsymbol{\nabla}\times\tilde{\mathbf{f}}_{{\mu}}\left(\mathbf{r}\right)-\left(\dfrac{\tilde{\omega}_{{\mu}}}{c}\right)^{2}
\epsilon(\mathbf{r},\tilde{\omega}_{\mu})\,\tilde{\mathbf{f}}_{{\mu}}\left(\mathbf{r}\right)=0,
\end{equation}
together with the Silver-M\"uller radiation condition~\cite{Kristensen2015} (open boundary conditions).
As a consequence of radiation condition, the QNM eigenfrequencies $\tilde{\omega}_{\mu}=\omega_{\mu}-i\gamma_{\mu}$ are complex, with a corresponding quality factor $Q_\mu=\omega_{\mu}/(2\gamma_{\mu})$, which also indicates that the radiation and/or absorption losses of the open systems are naturally included in the QNM theory. Indeed, QNMs can also be defined and computed for resonators with gain~\cite{PhysRevX.11.041020}, though a modified Fermi's golden rule is needed to compute the SE decay~\cite{franke_fermis_2021}.
%instead of being added phenomenologically in the normal mode (NM) theory.

The Green function includes 
the full linear response of an embedded dipole,
and is defined from
%nearly all response information of the %considering system.
% Putting a point dipole source in a system with relative permittivity $\epsilon(\mathbf{r},{\omega})$, then the Green function could be obtained through,
\begin{equation}\label{eq: Green_eq}
\boldsymbol{\nabla}\times\boldsymbol{\nabla}\times{\mathbf{G}}\left(\mathbf{r},\mathbf{r}^{\prime},\omega\right)-k_{0}^{2}
\epsilon(\mathbf{r},{\omega})\,{\mathbf{G}}\left(\mathbf{r},\mathbf{r}^{\prime},\omega\right)=k_{0}^{2}\delta(\mathbf{r}-\mathbf{r}^{\prime})\mathbf{I},
\end{equation}
where $\epsilon(\mathbf{r},{\omega})$
is the relative permittivity
or dielectric constant.
In general, analytical Green functions are only available for simple system with high degree of symmetry, so these generally have to be obtained numerically. However, one can obtain the
Green function for spatial points within or close to the resonator, using a highly accurate QNM expansion technique~\cite{leung_completeness_1994,ge_quasinormal_2014} 
\begin{equation}
\mathbf{G}\left(\mathbf{r},\mathbf{r}^{\prime},\omega\right)= \sum_{\mu} A_{\mu}\left(\omega\right)\,\tilde{\mathbf{f}}_{\mu}\left({\bf r}\right)\tilde{\mathbf{f}}_{\mu}\left({\bf r}^{\prime}\right),\label{eq: GFwithSUM_QNMs}
\end{equation}
where $A_{\mu}(\omega)={\omega}/[{2(\tilde{\omega}_{\mu}-\omega)}]$.

Using a classical light-matter theory, 
the {\it generalized} Purcell factor for a dipole with dipole moment $\mathbf{d} =d_{0}\mathbf{n}_{\rm d}$ at $\mathbf{r}_{0}$, is written as~\cite{Anger2006,kristensen_modes_2014}
\begin{align}\label{eq: QNMpurcell}
 \begin{split}
     &F_{{\rm P}}^{\rm }(\mathbf{r}_{0},\omega) =1+\frac{\Gamma(\mathbf{r}_{0},\omega)}{\Gamma_{0}(\mathbf{r}_{0},\omega)},
\end{split}
\end{align}
where the total SE rate is related with the total Green function in the frequency of interest, i.e., $\Gamma_{\rm }(\mathbf{r}_{0},\omega)=2\mathbf{d}\cdot{\rm Im}\{\mathbf{G}(\mathbf{r}_{0},\mathbf{r}_{0},\omega)\}\cdot\mathbf{d}/(\hbar\epsilon_{0})$,
where we assume ${\bf d}$ is real.
The background SE rate is  $\Gamma_{0}(\mathbf{r}_{0},\omega)=2\mathbf{d}\cdot{\rm Im}\{\mathbf{G}_{\rm B}(\mathbf{r}_{0},\mathbf{r}_{0},\omega)\} \cdot {\bf d}/(\hbar\epsilon_{0})$,
which is related with the homogeneous background Green function ${\bf G}_{\rm B}$; for a 2D TE dipole considered in the work, it is ${\rm Im}\{\mathbf{G}_{\rm B}(\mathbf{r}_{0},\mathbf{r}_{0},\omega)\}={\omega^2}/({8c^2})$.
Note that the additional $1$ originates from the background contribution for dipoles outside the resonator, and can be derived rigorously from a Dyson equation~\cite{ge_quasinormal_2014}.

To better understand the features of chiral radiation power flow, below we show the corresponding analytical expression from the known QNMs.
The scattered field (in real frequency space) at any point ${\bf r}$
from a dipole at ${\bf r}_{\rm d}$ is simply (using Eq.~\eqref{eq: GFwithSUM_QNMs}):
\begin{align}
\mathbf{E}^{\rm scatt}_{\rm QNMs}(\mathbf{r}_{\rm },\omega) &=\mathbf{G}(\mathbf{r}_{\rm },\mathbf{r}_{\rm d},\omega)\cdot\frac{\mathbf{d}}{\epsilon_{0}}\nonumber \\
=\frac{1}{\epsilon_{0}}\big[A_{1}(\omega)&\tilde{\bf f}_1({\bf r}) \tilde{\bf f}_1({\bf r}_{\rm d})
+ A_{2}(\omega)\tilde{\bf f}_2({\bf r}) \tilde{\bf f}_2({\bf r}_{\rm d})\big]
\cdot d_{0}{\bf n_{\rm d}},
\end{align}
and we will consider both
$\hat{\bf s}$ dipoles
and $\hat{\boldsymbol{\phi}}$ dipoles (i.e., ${\bf n}_{\rm d}=\hat{\bf s}$ or ${\bf n}_{\rm d}=\hat{\boldsymbol{\phi}}$). 
%Note that in an $x-y$ coordinate system, then $\hat{s}=\hat{x}\cos{\phi}+\hat{y}\sin{\phi}$ and $\hat{\phi}=-\hat{x}\sin{\phi}+\hat{y}\cos{\phi}$.

The corresponding magnetic scattered field is  obtained from
\begin{equation}
\begin{split}
\nabla\times\mathbf{E}^{\rm scatt}_{\rm QNMs}(\mathbf{r}_{\rm },\omega)
%&=-\mu_{0}\frac{\partial\mathbf{H^{\rm scatt}}(\mathbf{r}_{\rm },\omega)}{\partial t}\\
&=i\omega\mu_{0}\mathbf{H^{\rm scatt}_{\rm QNMs}}(\mathbf{r}_{\rm },\omega),
\end{split}
\end{equation}
%\red{SF: This is in frequency space, so it should be 
%\begin{equation}
%\nabla\times\mathbf{E}^{\rm scatt}(\mathbf{r}_{\rm },\omega)=i\omega\mu_{0}\mathbf{H^{\rm scatt}}(\mathbf{r}_{\rm },\omega),
%\end{equation}
%}
and  the corresponding Poynting vector is 
%\sh{update notation to match figure and also define the $\phi$ projected one shown in the graph}
\begin{equation}
\begin{split}\label{Eq:Spoyn}
\mathbf{S}^{\rm Poyn}_{\rm QNMs}(\mathbf{r}_{\rm },\omega)&=\frac{1}{2}{\rm Re}\big[\mathbf{E}^{\rm scatt}_{\rm QNMs}(\mathbf{r}_{\rm },\omega)\times\mathbf{H}^{\rm scatt~*}_{\rm QNMs}(\mathbf{r}_{\rm },\omega)\big]\\
&={S}_{x}^{\rm QNMs}(\mathbf{r}_{\rm },\omega)\hat{\bf x}+{S}_{y}^{\rm QNMs}(\mathbf{r}_{\rm },\omega)\hat{\bf y}\\
&={S}_{s}^{\rm QNMs}(\mathbf{r}_{\rm },\omega)\hat{\bf s}+{S}_{\phi}^{\rm QNMs}(\mathbf{r}_{\rm },\omega)\hat{\boldsymbol{\phi}},
\end{split}
\end{equation}
where $S_{x,y,s,\phi}^{\rm QNMs}(\mathbf{r}_{\rm },\omega)$ are the projections along $\hat{x}$,$\hat{y}$,$\hat{s}$,$\hat{\boldsymbol{\phi}}$. 

Thus, one will then find $S_{\phi}^{\rm QNMs}(\mathbf{r}_{\rm },\omega)<0$ ($>0$), corresponding to net chiral power flow  along the CW (CCW) direction. 
Using only the QNMs, we stress that we can compute the direction of the power flow and scattered fields analytically, for any dipole position and orientation. This further demonstrates the power of having an analytical expression for the Green function in terms of only the QNM properties.

\subsection{Normal mode limit for the Green function expansion}\label{sec: NMs}
It is useful to also compare with
the more usual NM limit of a Green function expansion. Specifically one would use %the Green function would be
\begin{align}
\begin{split}
\mathbf{G}_{\rm NMs}\left(\mathbf{r},\mathbf{r}^{\prime},\omega\right)&= \sum_{\mu} A_{\mu}\left(\omega\right)\,{\mathbf{f}}_{\mu}\left({\bf r}\right){\mathbf{f}}_{\mu}^{*}\left({\bf r}^{\prime}\right)\\
&\approx \sum_{\mu} A_{\mu}\left(\omega\right)\,\tilde{\mathbf{f}}_{\mu}\left({\bf r}\right)\tilde{\mathbf{f}}_{\mu}^{*}\left({\bf r}^{\prime}\right),\label{eq: GFwithSUM_NMs}
\end{split}
\end{align}
where we assume the NMs ${\mathbf{f}}_{\mu}\approx\tilde{\mathbf{f}}_{\mu}$ and the {\it conjugated product} form is used, i.e., the phase is neglected, which is an approximation since the actual NMs are ill-defined for the open systems.
A more strict NM expansion would use
 real mode functions, but typical numerical mode solvers actually use the QNM, if implementing open boundary conditions.
 Notably at ${\bf r}={\bf r}'$, there is no role of the phase at all, which is a key difference with QNM theory, which uses
 the {\it unconjugated} product (cf. Eq.~\eqref{eq: GFwithSUM_QNMs}).
 
 Using a NM expansion, the scattered field is now 
\begin{align}
\mathbf{E}^{\rm scatt}_{\rm NMs}(\mathbf{r}_{\rm },\omega) &=\mathbf{G}_{\rm NMs}(\mathbf{r}_{\rm },\mathbf{r}_{\rm d},\omega)\cdot\frac{\mathbf{d}}{\epsilon_{0}} \nonumber \\
=\frac{1}{\epsilon_{0}}\big[A_{1}(\omega)&\tilde{\bf f}_1({\bf r}) \tilde{\bf f}_1^{*}({\bf r}_{\rm d})
+ A_{2}(\omega)\tilde{\bf f}_2({\bf r}) \tilde{\bf f}_2^{*}({\bf r}_{\rm d})\big]
\cdot d_{0}{\bf n_{\rm d}},
\end{align}
and
\begin{equation}
\begin{split}
\nabla\times\mathbf{E}^{\rm scatt}_{\rm NMs}(\mathbf{r}_{\rm },\omega)
%&=-\mu_{0}\frac{\partial\mathbf{H^{\rm scatt}}(\mathbf{r}_{\rm },\omega)}{\partial t}\\
&=i\omega\mu_{0}\mathbf{H^{\rm scatt}_{\rm NMs}}(\mathbf{r}_{\rm },\omega).
\end{split}
\end{equation}
Then the corresponding Poynting vector is
\begin{equation}
\begin{split}\label{Eq:Spoyn_NMs}
\mathbf{S}^{\rm Poyn}_{\rm NMs}(\mathbf{r}_{\rm },\omega)&=\frac{1}{2}{\rm Re}\big[\mathbf{E}^{\rm scatt}_{\rm NMs}(\mathbf{r}_{\rm },\omega)\times\mathbf{H}^{\rm scatt~*}_{\rm NMs}(\mathbf{r}_{\rm },\omega)\big]\\
&={S}_{x}^{\rm NMs}(\mathbf{r}_{\rm },\omega)\hat{\bf x}+{S}_{y}^{\rm NMs}(\mathbf{r}_{\rm },\omega)\hat{\bf y}\\
&={S}_{s}^{\rm NMs}(\mathbf{r}_{\rm },\omega)\hat{\bf s}+{S}_{\phi}^{\rm NMs}(\mathbf{r}_{\rm },\omega)\hat{\boldsymbol{\phi}},
\end{split}
\end{equation}
where again one will focus on the projection $S_{\phi}^{\rm NMs}(\mathbf{r}_{\rm },\omega)$ to find out if the power flow will go along the CW (CCW) direction.

\subsection{Full numerical dipole method for a quantitative numerical check with no mode approximations}\label{sec: full}

In order to check the validity of the few mode QNMs/NMs solutions, we also employed the full numerical dipole method in the commercial COMSOL Multiphysics software~\cite{comsol}, which give the solutions to the full Maxwell equations using a classical dipole source.

First, the numerical Purcell factors of a point dipole could be given as
\begin{equation}\label{eq: Purcellfulldipole}
    F_{\rm P}^{\rm num}(\mathbf{r}_{0},\omega)=\frac{\int_{ L_{\rm c}}\hat{\mathbf{n}}\cdot {\bf S}_{\rm dipole,total}(\mathbf{r},\omega)d{L_{\rm c}} }{\int_{ L_{\rm c}}\hat{\mathbf{n}}\cdot {\bf S}_{\rm dipole,background}(\mathbf{r},\omega)d{L_{\rm c}} },
\end{equation}
where $\hat{\bf n}$ is the normal direction (pointing outward) to a small circle $L_{\rm c}$ surrounding the dipole, and ${\bf S}_{\rm dipole,total}$ (${\bf S}_{\rm dipole,background}$) is the corresponding Poynting vectors with the resonators (without the resonators, i.e., the background medium).

Second, the net power flow with the full numerical dipole method could be obtained through the numerical scattered fields $\mathbf{E}^{\rm scatt}_{\rm full}(\mathbf{r}_{\rm },\omega)$ and $\mathbf{H}^{\rm scatt}_{\rm full}(\mathbf{r}_{\rm },\omega)$, which are obtained by deducting the background fields $\mathbf{E}^{\rm B}_{\rm full}(\mathbf{r}_{\rm },\omega)$ and $\mathbf{H}^{\rm B}_{\rm full}(\mathbf{r}_{\rm },\omega)$ (without the resonator) from the total fields $\mathbf{E}^{\rm total}_{\rm full}(\mathbf{r}_{\rm },\omega)$ and $\mathbf{H}^{\rm total}_{\rm full}(\mathbf{r}_{\rm },\omega)$ (with the resonator). This is usually performed at a real frequency point close to the resonance.
Thus, the Poynting vector from full numerical dipole method is 
\begin{equation}
\begin{split}\label{Eq:Spoyn_full}
\mathbf{S}^{\rm Poyn}_{\rm full}(\mathbf{r}_{\rm },\omega)&=\frac{1}{2}{\rm Re}\big[\mathbf{E}^{\rm scatt}_{\rm full}(\mathbf{r}_{\rm },\omega)\times\mathbf{H}^{\rm scatt~*}_{\rm full}(\mathbf{r}_{\rm },\omega)\big]\\
&={S}_{x}^{\rm full}(\mathbf{r}_{\rm },\omega)\hat{\bf x}+{S}_{y}^{\rm full}(\mathbf{r}_{\rm },\omega)\hat{\bf y}\\
&={S}_{s}^{\rm full}(\mathbf{r}_{\rm },\omega)\hat{\bf s}+{S}_{\phi}^{\rm full}(\mathbf{r}_{\rm },\omega)\hat{\boldsymbol{\phi}},
\end{split}
\end{equation}
where again we need to focus on $S_{\phi}^{\rm full}(\mathbf{r}_{\rm },\omega)$ and compare them with the QNMs/NMs solutions. 

Although the full numerical dipole method results are reliable and always used as benchmark, the disadvantage is clear:
they offer no physical insight into the underlying physics and one must repeat the calculations for every single dipole point of interest. Also, the QNMs are required in order to have a rigorous quantized QNM~\cite{franke_quantization_2019,franke_quantized_2020} description of such structures, which is ultimately a requirement for quantum optical applications. 

\section{Quasinormal modes for index-modulated ring resonators near exceptional points}\label{sec: resultsQNMs}

\begin{figure*}
    \centering
    \includegraphics[width=1.85\columnwidth]{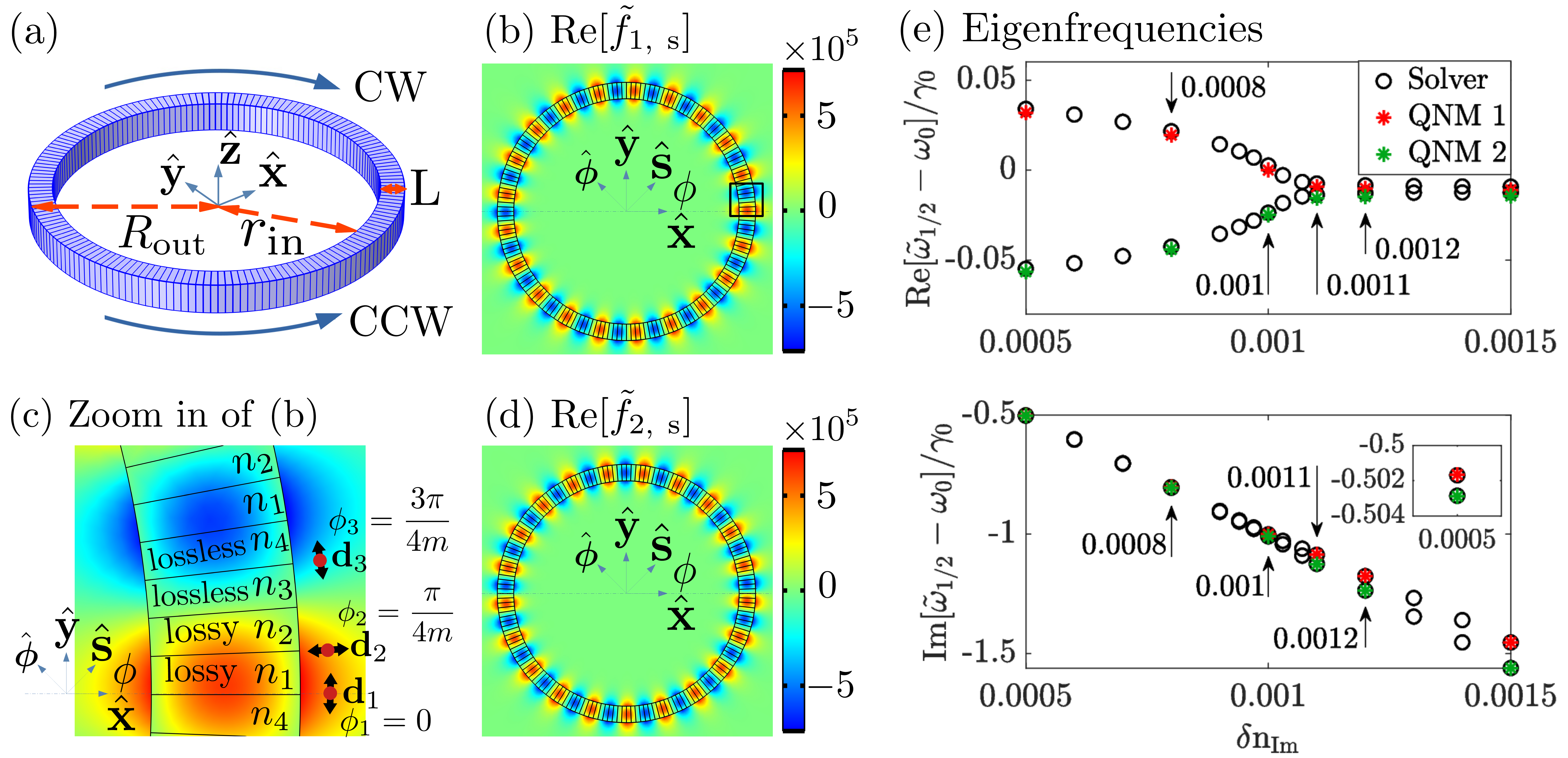}
       \caption{(a) 3D Schematic  of a refractive-index-modulated ring resonator, though in our calculations we use a 2D model, partly because of large memory requirements for the resonators. The two blue arrows label the direction of CW and CCW of the fields. 
       %The inner and outer radius (two red dashed arrows) of the ring are $r_{\rm in}=1109~$ nm and $R_{\rm out}=1271~$nm (width of the ring is ${\rm L}=R_{\rm out}-r_{\rm in}=162~$nm). In our calculations, we investigate a 2D ring resonator and couple it with an in-plane linearly polarized dipole. The ring is divided into $8m$~($m=21$) sections, where the refractive indices are $n_{1/2/3/4}$, which are distributed periodically along the CCW direction (see (c)). 
       Field distribution ${\rm Re}\big[\tilde{f}_s\big]$ (with units $m^{-1}$, s components) ($\tilde{\bf f}=\tilde{f}_{x}\hat{\bf x}+\tilde{f}_{y}\hat{\bf y}=\tilde{f}_{s}\hat{\bf s}+\tilde{f}_{\phi}\hat{\boldsymbol{\phi}}$) of (b) QNM~1 (${\rm Re}\big[\tilde{f}_{1,~s}\big]$) and (d) QNM~2 (${\rm Re}\big[\tilde{f}_{2,~s}\big]$) for ring structures with $\delta n_{\rm Re}=\delta n_{\rm Im}=0.001$. 
       (c) Zoom in of the black square in (b). 
       %In addition to general unit vectors $\hat{\bf x}$ and $\hat{\bf y}$, we also utilize $\hat{\bf s}=\hat{\bf x}\cos{\phi}+\hat{\bf y}\sin{\phi}$ and $\hat{\boldsymbol{\phi}}=-\hat{\bf x}\sin{\phi}+\hat{\bf y}\cos{\phi}$. Three example dipoles are shown, where one ($\mathbf{d}_{1}$) is a linearly polarized $\hat{\boldsymbol{\phi}}$ dipole at $\phi=\phi_{1}=0$ (i.e., $\hat{\bf y}$ dipole at $\phi=0$), one ($\mathbf{d}_{2}$) is a linearly polarized $\hat{\bf s}$ dipole at $\phi=\phi_{2}=\pi/4m$, and the left one ($\mathbf{d}_{3}$) is a linearly polarized $\hat{\boldsymbol{\phi}}$ dipole at $\phi=\phi_{2}=3\pi/4m$. 
       (e) Eigenfrequencies from QNMs ($\tilde{\omega}_{1/2}=\omega_{1/2}-i\gamma_{1/2}$, red and green stars) and approximate COMSOL solver (black circles) for the ring resonator as a function of $\delta n_{\rm Im}$ when keeping $\delta n_{\rm Re}=0.001$. The insert in the bottom figure shows the zoom in region close to $\delta n_{\rm Im}=0.0005$. One will find that there are two efficient QNMs even when $\delta n_{\rm Re}=\delta n_{\rm Im}=0.001$. 
       With the increase of $\delta n_{\rm Im}$, the decay rates  of these QNMs increase.
       Specifically, later we will focus on six cases, including $\delta n_{\rm Im}=0.0005,0.0015$ and four more cases that the arrows points to.
       %\sh{clarify that the schematic is 3d but we a 2d model}
       %. and compare the results from QNMs and the full numerical dipole simulation, where an excellent agreement is shown for the Purcell factors (cf.~Fig.~\ref{fig: purcell}). 
       %Note that the eigenfrequency of QNM~1 for $\delta n_{\rm Im}=0.001$ is $\omega_{0}-i\gamma_{0}\approx2.50543552\times10^{15} - i3.768337\times10^{11}$ (${\rm rad/s}$), i.e., $\omega_{0}$ and $\gamma_{0}$ are the real part and the opposite imaginary part of this pole eigenfrequency. 
       }\label{fig: scheeigen}
%\end{figure*}
%\begin{figure*}[th]
    \centering
    \includegraphics[width=0.65\columnwidth]{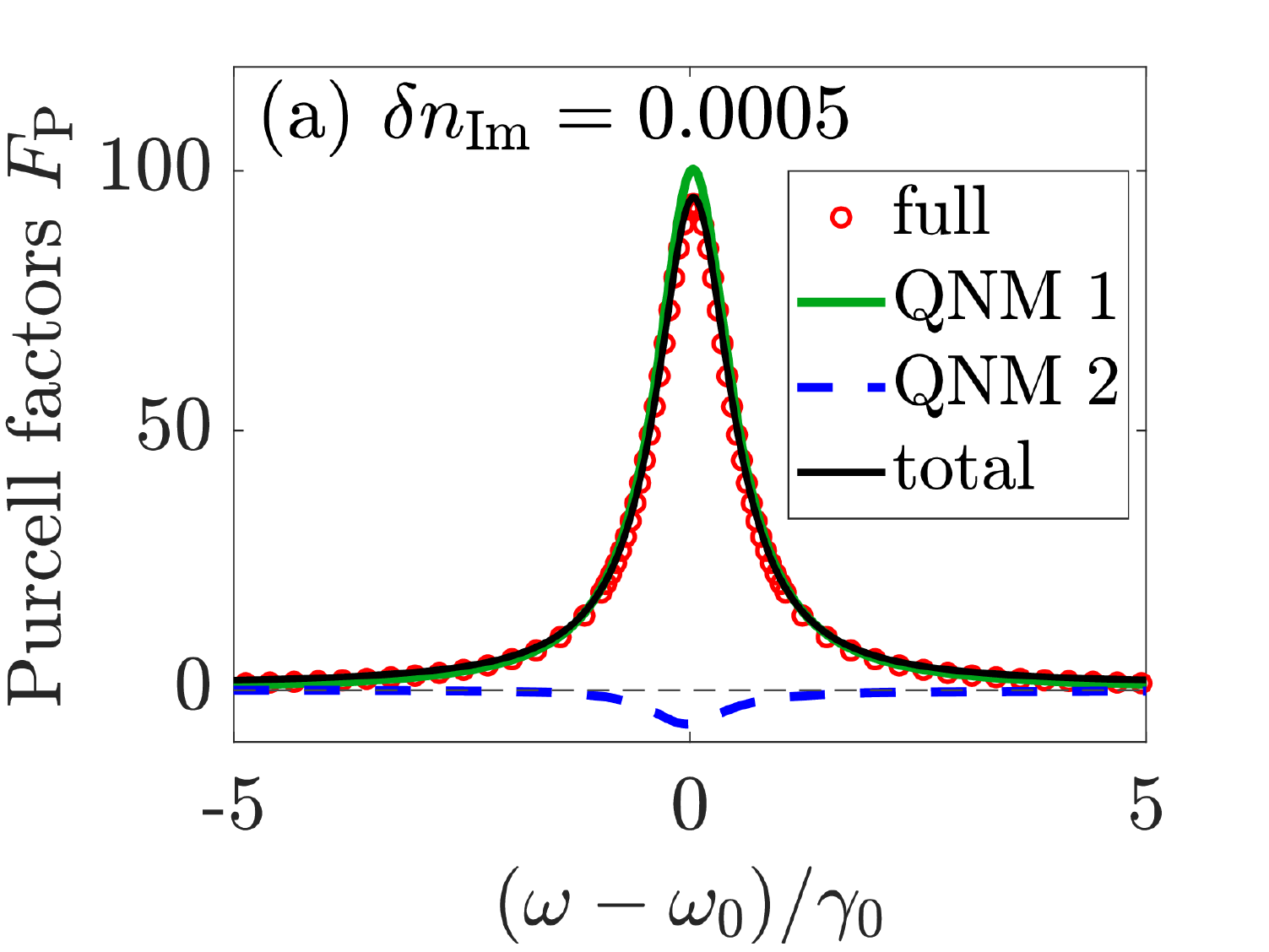}
    \includegraphics[width=0.65\columnwidth]{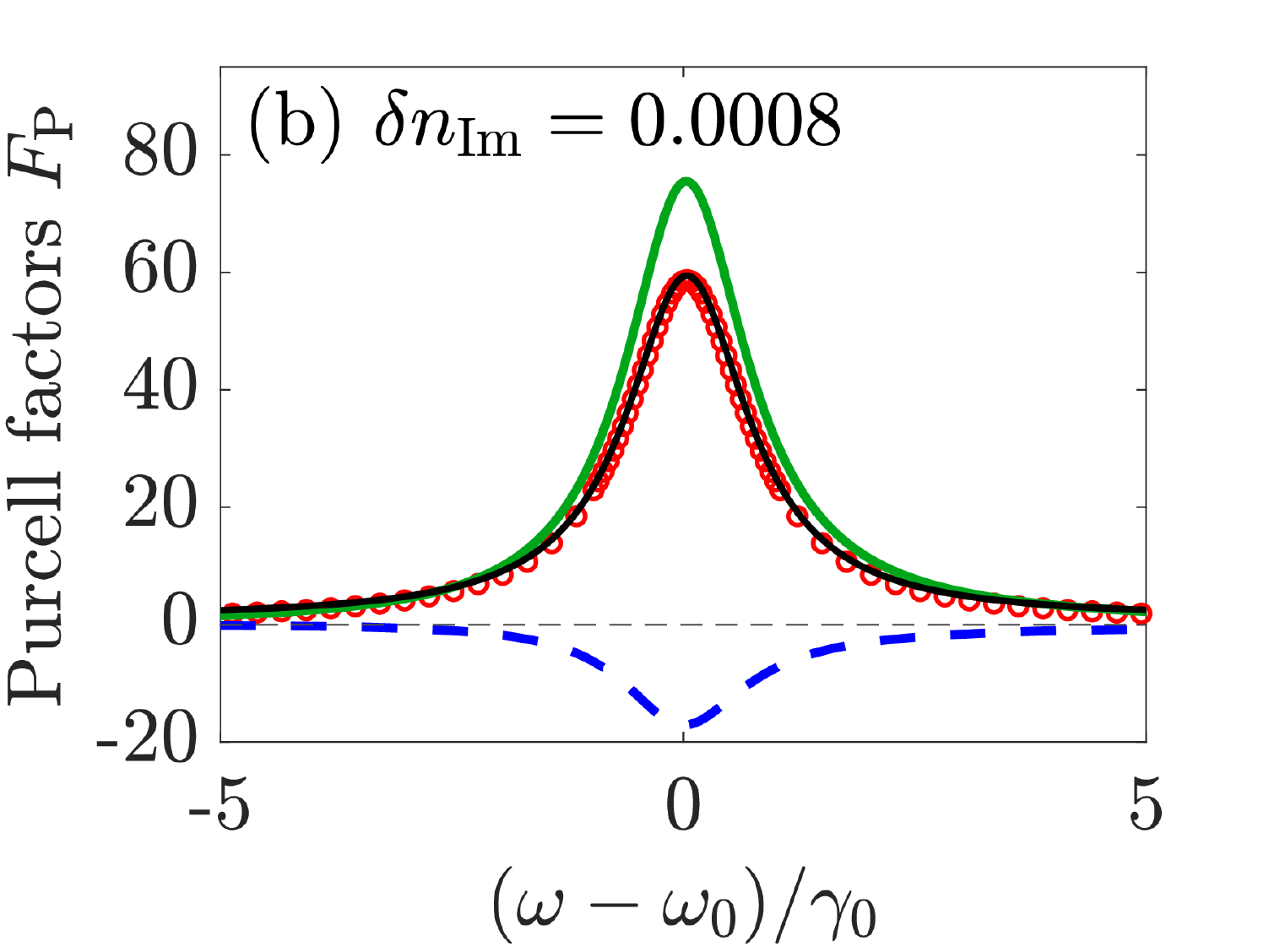}
    \includegraphics[width=0.65\columnwidth]{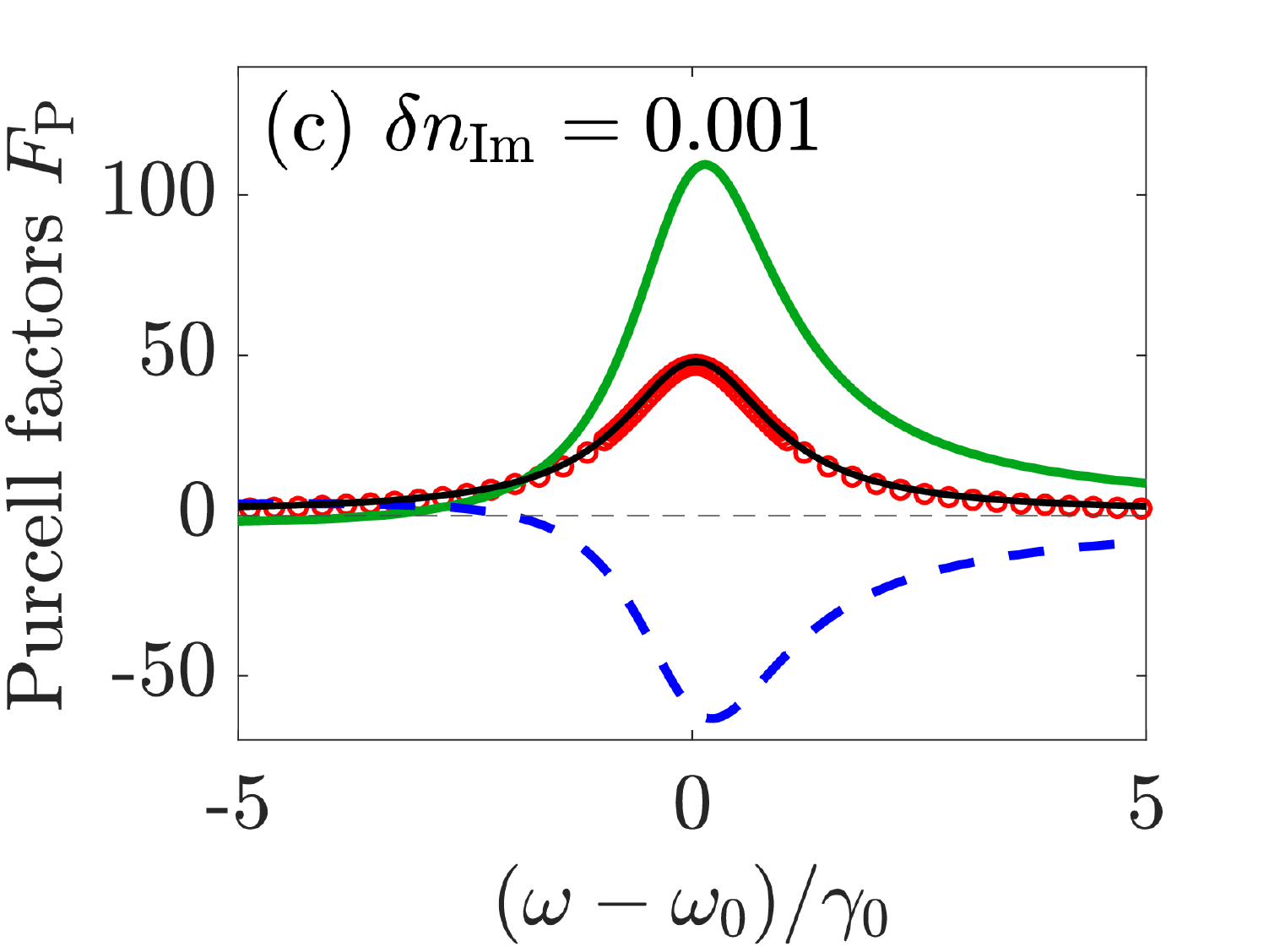}
    \includegraphics[width=0.65\columnwidth]{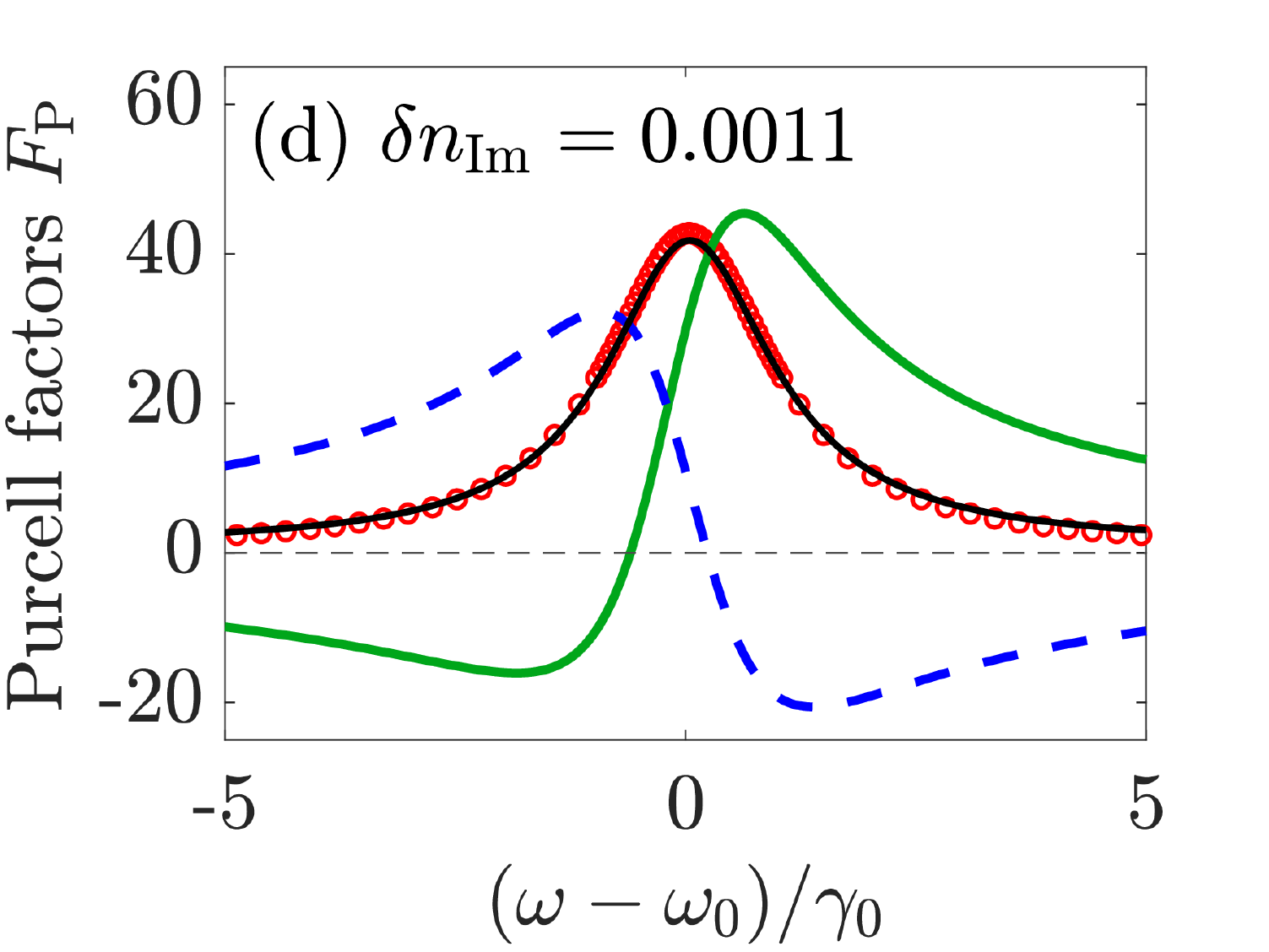}
    \includegraphics[width=0.65\columnwidth]{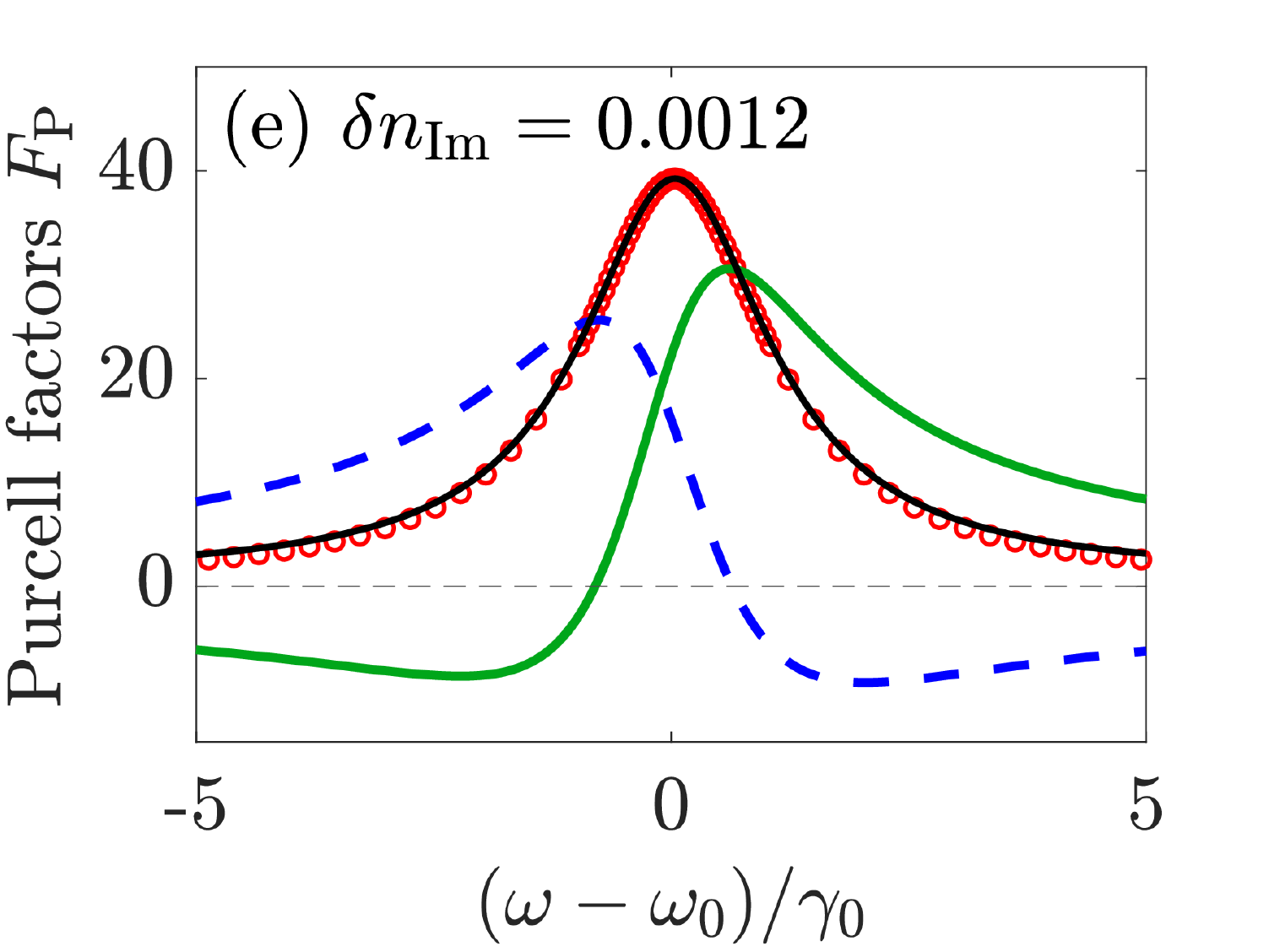}
    \includegraphics[width=0.65\columnwidth]{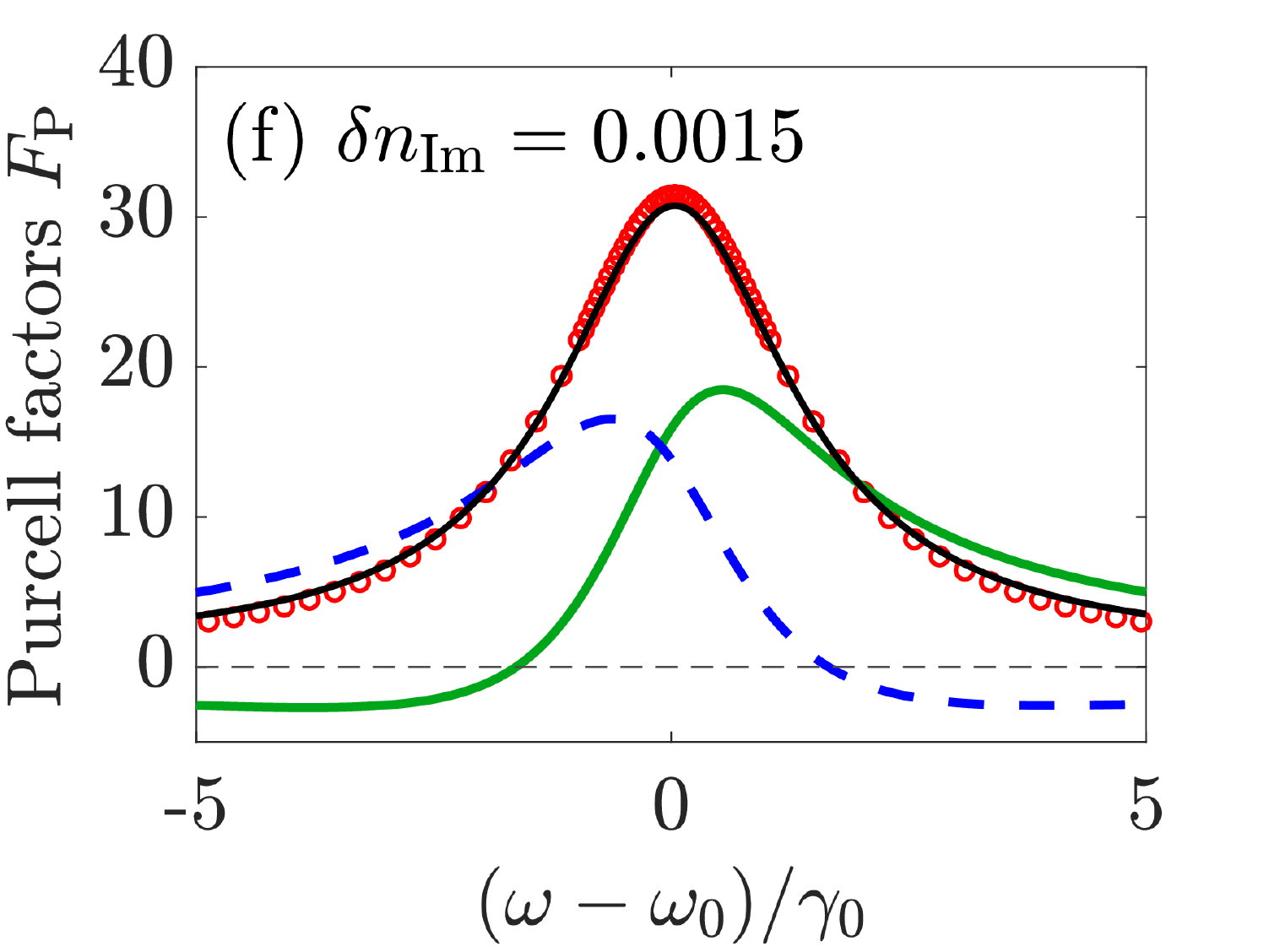}
    \caption{Purcell factors from QNMs (Eq.~\eqref{eq: QNMpurcell}) and full numerical dipole method (Eq.~\eqref{eq: Purcellfulldipole}) for six cases investigated with the analytical two QNM expansion (separate contributions from QNM~1 and QNM~2, and the total contribution), including (a) $\delta n_{\rm Im}=0.0005$, (b) $\delta n_{\rm Im}=0.0008$, (c) $\delta n_{\rm Im}=0.001$, (d) $\delta n_{\rm Im}=0.0011$, (e) $\delta n_{\rm Im}=0.0012$, and (f) $\delta n_{\rm Im}=0.0015$, while keeping $\delta n_{\rm Re}=0.001$. The Purcell factors are for a LP $\hat{\boldsymbol{\phi}}$ dipole at $\phi=0$ (i.e., a LP $\hat{\bf y}$ dipole at $\phi=0$) and $10~$nm away from the outer ring surface.
    $\omega_{0}$ and $\gamma_{0}$ are the real part and the opposite imaginary part of the pole eigenfrequency for QNM~1 with $\delta n_{\rm Im}=0.001$.
    %\sh{Describe what the legend means, and perhaps refer to the equation that was used to compute these to make it clear}
    %\sh{I would make the blue and green lines slighly thicker, and change the green to a derket green like on figs 5,6 }
    %Excellent agreement with a full numerical dipole method are obtained as with our coupled loss and gain resonators.
    %Their eigenfrequencies are marked in
    %Fig.~\ref{fig: scheeigen}. Although there are no gain regions in these resonators, we also note the appearance of modal negative Purcell factors, so the net sum is positive.% As mentioned before, $\omega_{0}$ and $\gamma_{0}$ are the real part and the opposite imaginary part of the eigenfrequency for QNM~1 with $\delta n_{\rm Im}=0.001$. The linewidth (associated with the opposite imaginary parts of the eigenfrequency) of these QNMs increases with $\delta n_{\rm Im}$, as also seen from Fig.~\ref{fig: scheeigen} (e).
   }\label{fig: purcell}
\end{figure*}

For numerical calculations, 
we consider a microring resonator similar to those in Refs.~\onlinecite{chen_revealing_2020,martin-cano_chiral_2019}, with a outer radius of $R_{\rm out}=1271~$nm and a inner radius of $r_{\rm in}=1109$~nm (width of $L=R_{\rm out}-r_{\rm in}=162$~nm) (cf.~Fig.~\ref{fig: scheeigen}(a)). 
The two blue arrows in Fig.~\ref{fig: scheeigen}(a) show the direction of CW and CCW fields. In our calculations, we  investigate a 2D ring resonator and couple it with an in-plane LP dipole. This geometry yields
a TE-like mode.

As shown in Fig.~\ref{fig: scheeigen}(c), the refractive index of the ring is modulated periodically (along the CCW direction), which has the following form~\cite{chen_revealing_2020,feng_single-mode_2014,miao_orbital_2016}:
 \begin{align}
 \begin{split}\label{eq: period_n}
    n_{1}&=n_{0}+\delta n_{\rm Re}+i\delta n_{\rm Im},~\big[l\pi/m<\phi<(l+1/4)\pi/m\big], \\
    n_{2}&=n_{0}+i\delta n_{\rm Im},~\big[(l+1/4)\pi/m<\phi<(l+1/2)\pi/m\big],\\
    n_{3}&=n_{0},~\big[(l+1/2)\pi/m<\phi<(l+3/4)\pi/m\big],\\
    n_{4}&=n_{0}+\delta n_{\rm Re},~\big[(l+3/4)\pi/m<\phi<(l+1)\pi/m\big],
\end{split}
 \end{align}
where $l=0,1,2, \dots, 2m-1$, and $m$ is the azimuthal mode number. There are $2m$ periods and $8m$ sections in total.
Notably, such modulation only works for the mode with the specific azimuthal mode number $m$ (radial mode number is usually set as $q=1$, but the mode with $q=2$ etc. also has similar trends as $q=1$). In other words, one could design the modulation for the particular mode of interest. 
Here, we choose a TE mode with $q=1$, and $m=21$, whose resonance is around $752$~nm (visible spectrum).
%, that is to say, the modulations above is only working for our chosen mode \sh{can we say this for sure, or is it just an example, and is it a similar mode to the one in the Nature paper? - brief paragraph on this would be useful to add if we know the answers to help guide the reader and also to say how general these findings are}.
In the following, we will keep $n_{0}=3.0$ and $\delta n_{\rm Re}=0.001$, while $\delta n_{\rm Im}$ is changed in the range of $0.0005\sim0.0015$ (positive, lossy materials; the case with gain will also be investigated in Sec.~\ref{sec: discussion}). 

Using an efficient dipole-excitation QNM technique~\cite{bai_efficient_2013-1}, we employ a LP dipole, $\hat{\boldsymbol{\phi}}$, placed at $\phi=\phi_1=0$ (which is also a linear $\hat{\bf y}$ dipole at $\phi=0$) and $10~$nm away from the ring surface (the bottom dipole $\mathbf{d}_{1}=d_{0}\hat{\boldsymbol{\phi}}=d_{0}\hat{\bf y}$ in Fig.~\ref{fig: scheeigen}(c), schematically showing its position and polarization).
Two QNMs (which we label as QNM~1 and QNM~2) are found in the frequency regime of interest, whose eigenfrequencies ($\tilde{\omega}_{1}=\omega_{1}-i\gamma_{1}$ and $\tilde{\omega}_{2}=\omega_{2}-i\gamma_{2}$) are shown in Fig.~\ref{fig: scheeigen}(e) (red and green stars). We also investigate the eigenfrequecies from the approximate COMSOL direct eigenfrequency solver (black circles), which are generally highly accurate for high $Q$ resonators though it's not a robust nonlinear eigenmode solver.  
%\sh{I think this needs  more explanation here, or a general reader will be confused what you mean by approximate}, 
The results agree well with those obtained from the dipole-excited QNM technique (since mainly we are dealing with high $Q$ resonators, around $2100$ to $6600$ for various modulated index cases).
%\sh{give the $Q$ factors as well to make this clear}. 
%Note in this section, QNM~1 (QNM~2) and $\tilde{\omega}_{1}$ ($\tilde{\omega}_{1}$) are for coupled modes of the index-modulated ring resonators.
Specifically, the eigenfrequency of QNM~1 for $\delta n_{\rm Im}=0.001$ is $\omega_{0}-i\gamma_{0}\approx2.50543552\times10^{15} - i3.768337\times10^{11}$ (${\rm rad/s}$), i.e., $\omega_{0}$ and $\gamma_{0}$ are defined as the real part and the opposite imaginary part of this pole eigenfrequency.

The QNM field distributions ${\rm Re}\big[\tilde{f}_s\big]$ (s components),
%(with units ${\rm m}^{-1}$) 
with $\tilde{\bf f}=\tilde{f}_{x}\hat{\bf x}+\tilde{f}_{y}\hat{\bf y}=\tilde{f}_{s}\hat{\bf s}+\tilde{f}_{\phi}\hat{\boldsymbol{\phi}}$, of QNM~1 and QNM~2 for the ring structures, with $\delta n_{\rm Re}=\delta n_{\rm Im}=0.001$, are shown in Figs.~\ref{fig: scheeigen}(b) and (d). 
%Here we just show their real parts.
%, though their imaginary parts will also contribute to responses, such as Purcell factors. 
For clarity, the black square area in Fig.~\ref{fig: scheeigen}(b) is enlarged and shown in Fig.~\ref{fig: scheeigen}(c), where the detailed periodic refractive index is displayed. In addition to the general ($\hat{\bf x},\hat{\bf y}$) basis, we also utilize a polar ($\hat{\bf s},\hat{\boldsymbol{\phi}}$) basis. The corresponding unit vectors are related from $\hat{\bf s}=\hat{\bf x}\cos{\phi}+\hat{\bf y}\sin{\phi}$ and $\hat{\boldsymbol{\phi}}=-\hat{\bf x}\sin(\phi)+\hat{\bf y}\cos{\phi}$ (the positive direction of polar angle $\phi$ is along CCW direction). 

As mentioned before, with the dipole QNM technique, we use a LP $\hat{\boldsymbol{\phi}}$ dipole at $\phi=0$ (the bottom one $\mathbf{d}_{1}$ in Fig.~\ref{fig: scheeigen}(c)). Once we obtain the QNMs, then we can easily couple them to any in-plane LP dipoles with any polarization using a general Green function theory described in Sec.~\ref{sec: theory}.
%below \blue{SF: Do you mean 'described above', i.e. in Section II?}. 
We schematically show three example dipoles in Fig.~\ref{fig: scheeigen}(c), where one ($\mathbf{d}_{1}$) is a LP $\hat{\boldsymbol{\phi}}$ dipole at $\phi=\phi_{1}=0$ (i.e., $\hat{\bf y}$ dipole at $\phi=0$), one ($\mathbf{d}_{2}$) is a LP $\hat{\bf s}$ dipole at $\phi=\phi_{2}=\pi/4m$, and the left one ($\mathbf{d}_{3}$) is a LP $\hat{\boldsymbol{\phi}}$ dipole at $\phi=\phi_{2}=3\pi/4m$. For these dipole positions, we can model the emitted radiation analytically using the QNMs, and also from a direct solution of Maxwell's equations. To verify the accuracy of our QNM predictions for dipole emitted radiation, we will carry out both approaches below. In addition, we will show the essential role of the QNM phase, and point out the clear failure of using a NM expansion for the Green function response.

\section{Purcell factors versus frequency for different dipole locations}\label{sec: Purcellfactors}

To confirm the accuracy of our two QNM description for the
index-modulated ring resonators, we next calculate the Purcell factors analytically from the QNM properties (Eq.~\eqref{eq: QNMpurcell}) for a $\hat{\boldsymbol{\phi}}$ dipole at $\phi=0$ (equal to $\hat{\bf y}$ at $\phi=0$, $\mathbf{d}_{1}$ in Fig.~\ref{fig: scheeigen} (c)), and compare them with the full numerical dipole method (Eq.~\eqref{eq: Purcellfulldipole}). For all configurations investigated, an excellent agreement is shown as a function of frequency as demonstrated in Fig.~\ref{fig: purcell}, where the results for six cases with the analytical two QNM expansion are studied, including $\delta n_{\rm Im}=0.0005,0.0008,0.001,0.0011,0.0012,0.0015$ while keeping $\delta n_{\rm Re}=0.001$. 
Their eigenfrequencies are marked in Fig.~\ref{fig: scheeigen}(e). Although there are no material gain regions in these resonators, we also note the appearance of modal negative Purcell factors.
This level of agreement is unusual in the sense that
the eigenmodes in Ref.~\cite{chen_revealing_2020} were reported to be completely decoupled, which was used to argue  a chiral emission, an effect that is not expected nor obtained for a regular ring resonator. We will connect to the power flow emission below, and show that it is also well explained in terms of the underlying QNMs.

\section{Chiral power flow from linearly  polarized dipoles}\label{sec: chiralflow}

To directly connect with the experimentally measured chirality in Ref.~\onlinecite{chen_revealing_2020} from LP dipole emitters, here we show chiral power flow using only the QNM propagators in a similar microring with refractive index modulation.
We can  easily obtain the power flow by computing the
scattered fields and the Poynting vector from QNMs and the QNM propagator, and thus the calculations are basically instantaneous (see Eq.~\eqref{Eq:Spoyn}).

\begin{figure*}
    \centering
    \includegraphics[width=1.6\columnwidth]{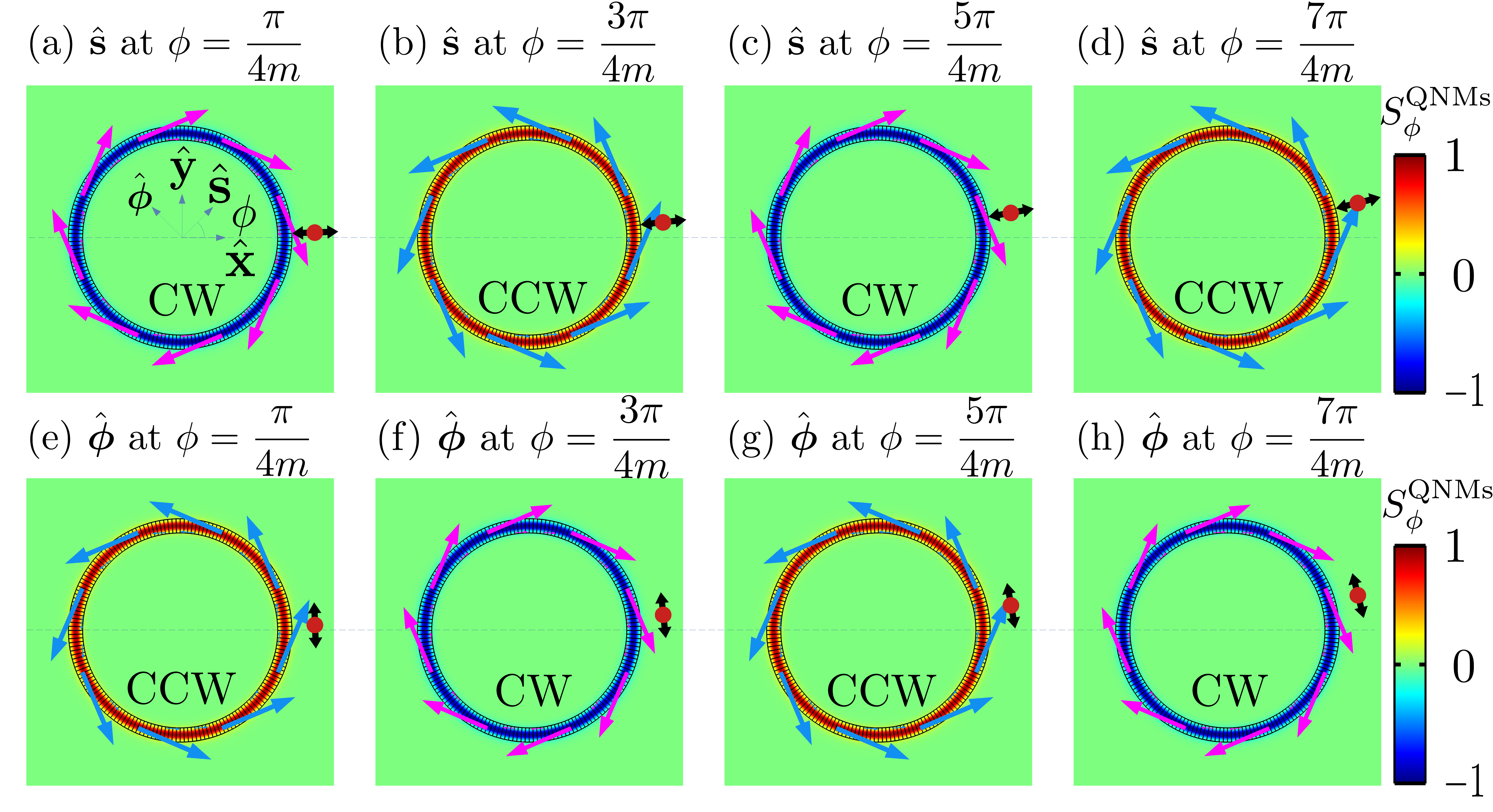}
       \caption{QNM calculations of power flow for various dipoles. We show the QNM Poynting vector ($\mathbf{S}^{\rm Poyn}_{\rm QNMs}=S_{x}^{\rm QNMs}\hat{\bf x}+S_y^{\rm QNMs}\hat{\bf{y}}=S_{s}^{\rm QNMs}\hat{\bf s}+S_{\phi}^{\rm QNMs}\hat{\boldsymbol{\phi}}$) using Eq.~\eqref{Eq:Spoyn}, for (a,b,c,d) $\hat{\bf s}$ and (e,f,g,h) $\hat{\boldsymbol{\phi}}$ dipoles at $\phi=\pi/4m$, $\phi=3\pi/4m$, $\phi=5\pi/4m$, and $\phi=7\pi/4m$ with $\delta n_{\rm Re}=\delta n_{\rm Im}=0.001$ (the radial distance between the dipole and the ring surface is $10~$nm). The red dot with double black arrows schematically describes the position and polarization of the LP dipoles. The Poynting vectors are labelled with magenta (for CW direction) and blue (for CCW direction) arrows. The distributions show the projection $S^{\rm QNMs}_{\phi}$ along $\hat{\boldsymbol{\phi}}$. These power flows are calculated at an example real frequency of $\omega=2.50543\times10^{15}$~(${\rm rad/s}$) (close to QNM resonance). In the ring region, one will find $S^{\rm QNMs}_{\phi}<0$ (it is shown in blue for the entire ring) for CW propagation and $S^{\rm QNMs}_{\phi}>0$ (red distribution in the entire ring) for CCW propagation.}\label{fig: QNMflow}
\end{figure*}

\begin{figure*}
    \centering
    \includegraphics[width=1.6\columnwidth]{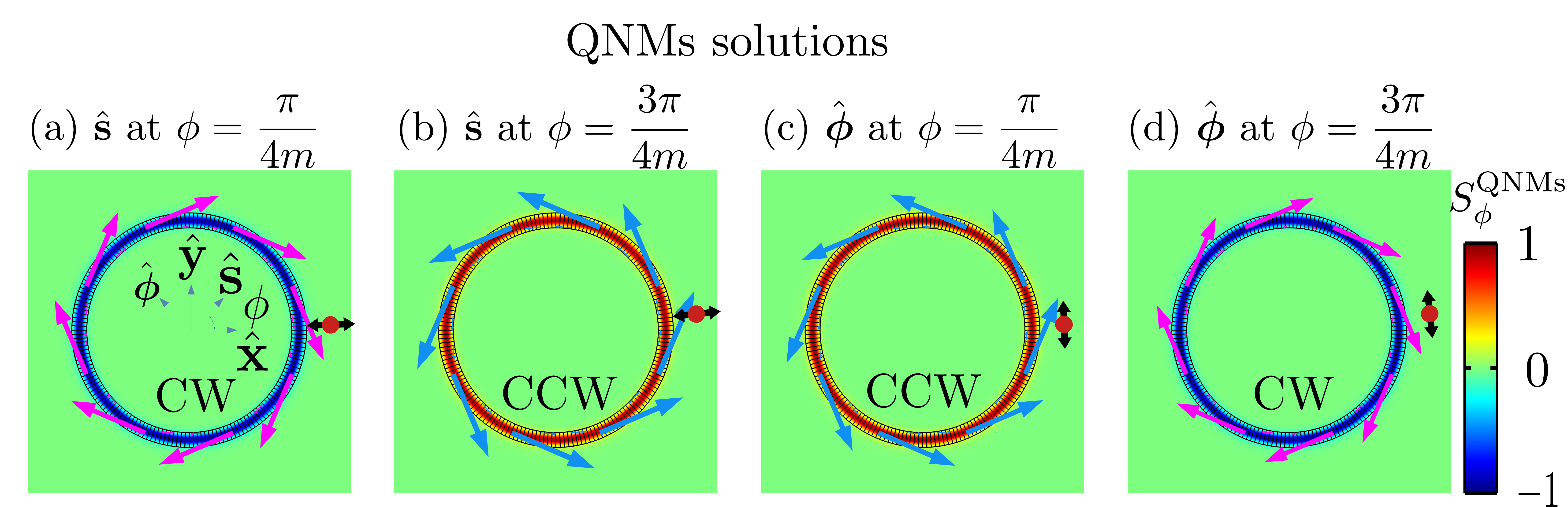}
    \includegraphics[width=1.6\columnwidth]{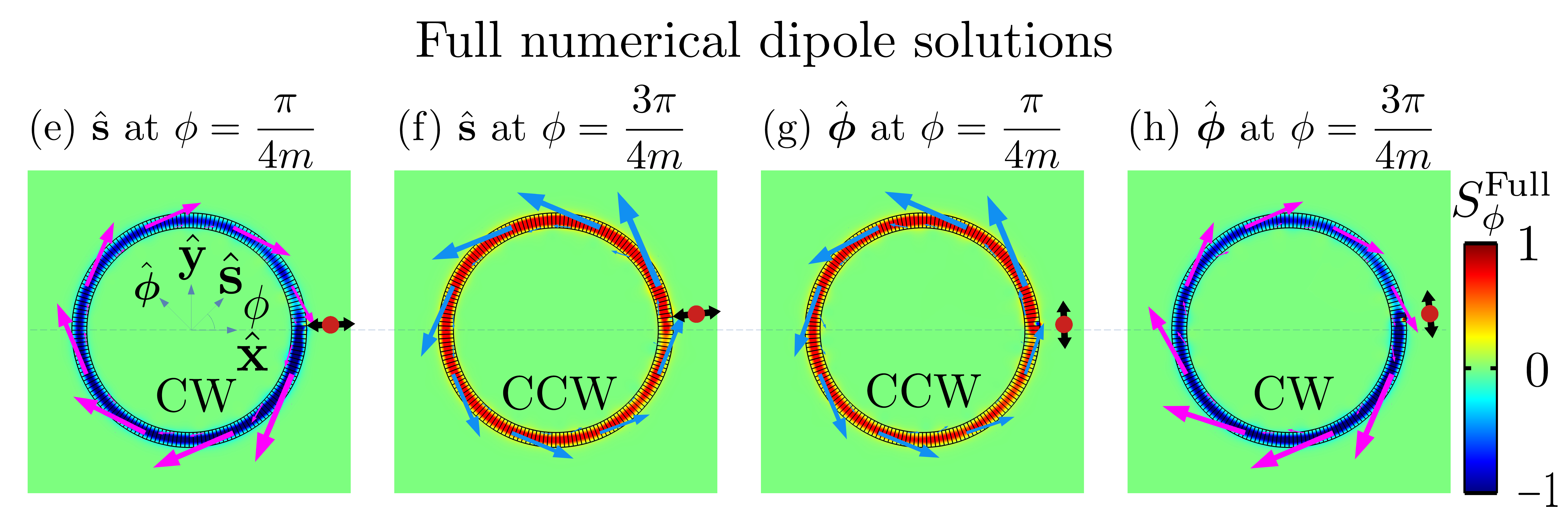}
    \includegraphics[width=1.6\columnwidth]{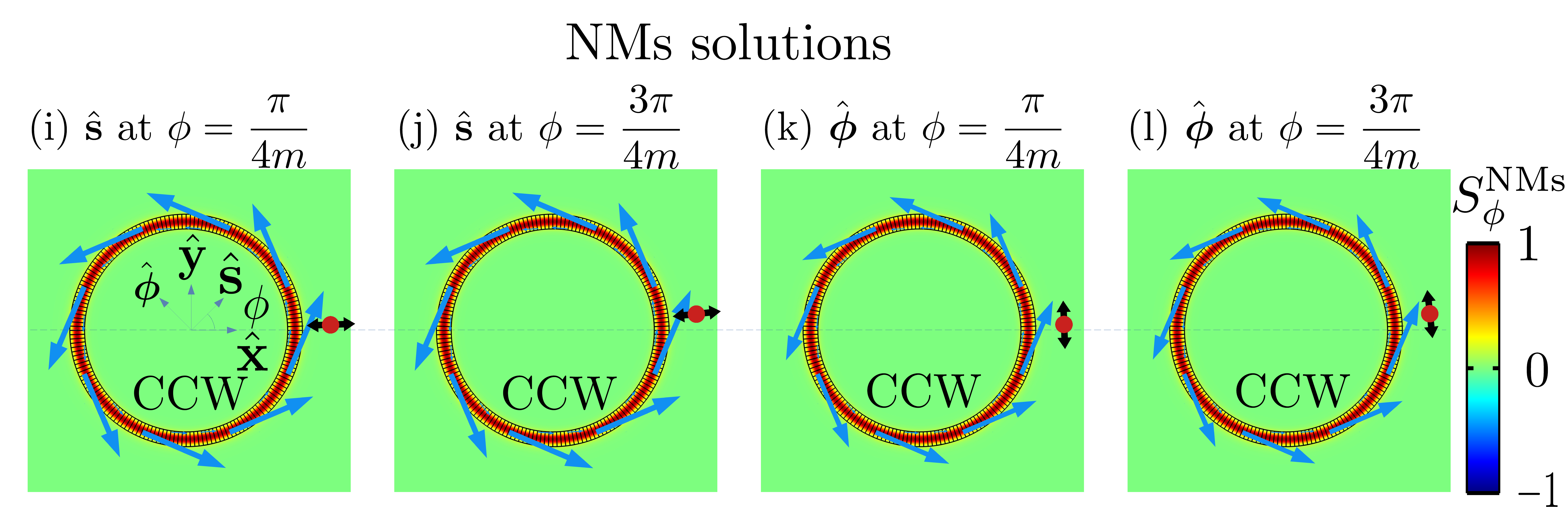}
      \caption{Power flow from (a-d) QNMs solutions (Eq.~\eqref{Eq:Spoyn}), (e-h) full numerical dipole solutions (Eq.~\eqref{Eq:Spoyn_full}) and (i-l) NMs solutions (Eq.~\eqref{Eq:Spoyn_NMs}) for $\hat{\bf s}$ or $\hat{\boldsymbol{\phi}}$ at $\phi=\pi/4m$ or $\phi=3\pi/4m$. QNMs solutions recover the full numerical dipole results perfectly while the NM solutions fail in  general (always CCW).  
      }\label{fig: compare}
\end{figure*}

In this section, we focus on the case with $\delta n_{\rm Re}=\delta n_{\rm Im}=0.001$, which is closest to the EP as seen from Fig.~\ref{fig: scheeigen}(e) and Fig.~\ref{fig: purcell}(c).
Figures \ref{fig: QNMflow}(a-h) shows the chiral power flow (Eq.~\eqref{Eq:Spoyn}) from our QNM model, for $\hat{\bf s}$ and $\hat{\boldsymbol{\phi}}$ dipole at several specific positions, including $\phi=\pi/4m$ (center of two lossy sections, see Fig.~\ref{fig: scheeigen}(c)), $\phi=3\pi/4m$ (center of two lossless sections), $\phi=5\pi/4m$ (center of two lossy sections), and $\phi=7\pi/4m$ (center of two lossless sections) (the radial distance between the dipole and the ring surface is $10~$nm). The red dot with double black arrows in Fig.~\ref{fig: QNMflow} schematically shows the position and polarization of these LP dipoles (though this is not to scale). Also note that these power flows presented in Fig.~\ref{fig: QNMflow} are calculated at an examplary real frequency $\omega=2.50543\times10^{15}$~(${\rm rad/s}$) (close to QNM resonance), and we stress that the chirality will remain the same in a range around $\omega_{0}\pm1.5\gamma_{0}$.

The first thing to observe, is that when a LP $\hat{\bf s}$-dipole is placed at $\phi=\pi/4m$ (Fig.~\ref{fig: QNMflow} (a)), the net power flow ($\mathbf{S}_{\rm QNMs}^{\rm Poyn}$, Eq.~\eqref{Eq:Spoyn}), goes along the CW direction (indicated by the magenta arrows). The  distribution shown is the projection $S_{\phi}^{\rm QNMs}$ along $\hat{\boldsymbol{\phi}}$. One will find that $S_{\phi}^{\rm QNMs}<0$ in the ring region, which further confirms that the net energy flow goes along the CW direction.
This also supports the experimental findings in Ref.~\onlinecite{chen_revealing_2020}, where they measured CW radiation when putting a LP dipole at $\phi=\pi/4l$ (center of lossy region) (azimuthal mode number $l=1$ in their considered structure, and we are using azimuthal mode number $m=21$ for a similar ring structure). However, our interpretation is drastically different. Instead of arguing with the help of a decoupling from the eigenmodes and coupling to a missing dimension, we find that our net chiral power flows are fully explained from the underlying QNMs, which are the correct natural eigenmodes of such resonators\footnote{Assuming that one is not at a perfect EP, which we have discussed earlier is highly unlikely, and thus not a practical concern. }.

In addition, the azimuthal period of such phenomenon is $\pi/m$, i.e., when a LP $\hat{\bf s}$ dipole is located at $\phi=\frac{\pi}{4m}+p\frac{\pi}{m}$ (center of the two lossy regions), where $p=0,1,2,...2m-1$, the net power flow will go along the CW direction, such as at $\phi=5\pi/4m$ shown in Fig.~\ref{fig: QNMflow} (c). 

Moreover, we also find that such unusual properties are not limited to $\hat{\bf s}$ dipoles at $\phi=\frac{\pi}{4m}+p\frac{\pi}{m}$~($p=0,1,2,...2m-1$). As shown in Fig.~\ref{fig: QNMflow} (b,d), when a LP $\hat{\bf s}$ dipole is placed at $\phi=\frac{3\pi}{4m}+p\frac{\pi}{m}$ (center of the two lossless sections, where $p=0,1,2,...2m-1$), the net power flow is along the CCW direction, confirmed with both Poynting vectors $\mathbf{S}_{\rm QNMs}^{\rm Poyn}$ (Eq.~\eqref{Eq:Spoyn}) (blue arrows) and projection $S_{\phi}^{\rm QNMs}>0$.

Furthermore, we found that such special chiral radiation are not only found with $\hat{\bf s}$ dipoles but also with LP $\hat{\boldsymbol{\phi}}$ dipole. As shown in Fig.~\ref{fig: QNMflow} (e-h), when a LP $\hat{\boldsymbol{\phi}}$ dipole is placed at $\phi=\frac{\pi}{4m}+p\frac{\pi}{m}$ ($\phi=\frac{3\pi}{4m}+p\frac{\pi}{m}$, $p=0,1,2,...2m-1$), the net power flow will go along the CCW (CW) direction, again confirmed with Poynting vectors $\mathbf{S}_{\rm QNMs}^{\rm Poyn}$ with blue arrows (magenta arrows) and projection $S_{\phi}^{\rm QNMs}>0$ ($<0$).

To further justify that our QNM picture is both correct
and rigorously accurate for describing this chiral emission,
we have confirmed these unusual properties with full numerical dipole calculations as well as NM solutions. 
For a comprehensive comparison, we will focus on four cases, i.e., $\hat{\bf s}$ or $\hat{\boldsymbol{\phi}}$ dipoles placed at $\phi=\pi/4m$ or $\phi=3\pi/4m$, and all other special cases could be obtained automatically due to the periodic modulated index.
For convenience, we show the QNM solutions given in Fig.~\ref{fig: QNMflow} (a-b) and (e-f) again in Fig.~\ref{fig: compare} (a-d).
The corresponding full numerical dipole solutions (Eq.~\eqref{Eq:Spoyn_full}) are shown in Fig.~\ref{fig: compare} (e-h), and one will find excellent agreement between QNM solutions and full numerical dipole solutions.

It is also insightful to compare with a 
NM solution, to assess the role of the QNM phase.
In contrast to the QNM approach, the NM model (Eq.~\eqref{Eq:Spoyn_NMs}) fails to capture the crucial effect of the change of power flow direction. Indeed, as shown in Fig.~\ref{fig: compare} (i-l), the flow direction always stays in CCW direction no matter where the dipole is placed and how it it polarized. To investigate this in more detail, we inspect the analytical structure of the power flow formulas. We recognize, that the crucial difference appears in the off-diagonal elements of the Poynting vector. To see this more clearly, we rewrite the Poynting vectors as $\mathbf{S}^{\rm Poyn}_{\rm NMs}=\mathbf{S}^{\rm Poyn}_{\rm ref}+\sin(\theta_1-\theta_2)\mathbf{S}^{\rm Poyn}_{\rm diff}$ and $\mathbf{S}^{\rm Poyn}_{\rm QNMs}=\mathbf{S}^{\rm Poyn}_{\rm ref}-\sin(\theta_1-\theta_2)\mathbf{S}^{\rm Poyn}_{\rm diff}$,
where $\theta_1,\theta_2$ are the complex phases of the dipole-projected QNM eigenfunctions 1,2 ($\tilde{\mathbf{f}}_{1/2}(\mathbf{r}_{d})\cdot\mathbf{n}_{\rm d}=\big|\tilde{\mathbf{f}}_{1/2}(\mathbf{r}_{d})\cdot\mathbf{n}_{\rm d}\big|e^{i \theta_{1/2}}=\big|\tilde{f}_{1/2,\rm d}(\mathbf{r}_{\rm d})\big|e^{i \theta_{1/2}}$) and
\begin{align}
    \mathbf{S}^{\rm Poyn}_{\rm diff}&(\mathbf{r}_{\rm },\omega)\nonumber\\
    =&\big|\tilde{f}_{1,\rm d}(\mathbf{r}_{\rm d})\big|\big|\tilde{f}_{2,\rm d}(\mathbf{r}_{\rm d})\big|{\rm Re}\big[A_1(\omega)A_2^*(\omega)\tilde{\mathbf{s}}_{12}(\mathbf{r},\omega)\big]\nonumber\\
    &-\big|\tilde{f}_{1,\rm d}(\mathbf{r}_{\rm d})\big|\big|\tilde{f}_{2,\rm d}(\mathbf{r}_{\rm d})\big|{\rm Re}\big[A_2(\omega)A_1^*(\omega)\tilde{\mathbf{s}}_{21}(\mathbf{r},\omega)\big],
\end{align}
with $\tilde{\mathbf{s}}_{12}(\mathbf{r},\omega)=\tilde{\mathbf{f}}_1(\mathbf{r})\times\big[\boldsymbol{\nabla}\times\tilde{\mathbf{f}}_2^*(\mathbf{r})\big]/(2\mu_0\omega)$ and  $\tilde{\mathbf{s}}_{21}(\mathbf{r},\omega)=\tilde{\mathbf{f}}_2(\mathbf{r})\times\big[\boldsymbol{\nabla}\times\tilde{\mathbf{f}}_1^*(\mathbf{r})\big]/(2\mu_0\omega)$.

The sign difference between the QNM and NM model is induced by use of the unconjugated and conjugated form, respectively, and depends on the difference of the QNM phases, which are highly position-dependent. In that sense, even if we take full account of the QNM eigenfrequency and eigenfunctions for the NM model, it does not recover the crucial change of the power flow direction, and applying the usual NM assumptions would lead to even more differences.   %Note, that the diagonal elements of the Poynting vector are not phase dependent. 

\section{Discussion on frequency dependent chiral flow and chiral flow with gain media}\label{sec: discussion}

\begin{figure*}
    \centering
    \includegraphics[width=1.8\columnwidth]{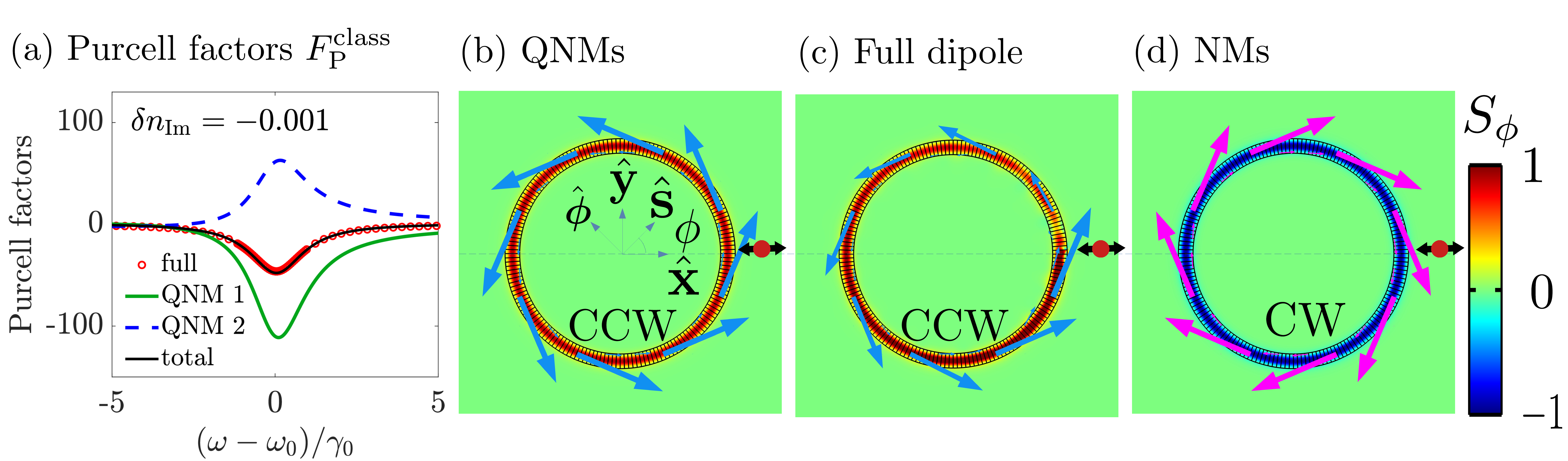}
      \caption{Ring resonators with gain. (a) Classical Purcell factors (Eq.~\eqref{eq: QNMpurcell}, here we refer it as $F_{\rm P}^{\rm class}$) from the QNMs. The power flow of a LP $\hat{\bf s}$ dipole at $\phi=\pi/4m$ from (b) QNMs solutions, (c) full numerical dipole solutions and (d) NMs solutions for ring resonator with gain ($\delta n_{\rm Im}=-0.001$, $\delta n_{\rm Re}=0.001$). 
      }\label{fig: gain}
\end{figure*}

Apart from above discussed position and polarization-dependent chiral radiation, we also observe, that the chirality is sensitive to a change in frequency at special locations. 
For example, for $\hat{\bf s}$ dipoles at $\phi=p\frac{\pi}{m}$ ($p=0,1,2,...2m-1$)
%(center of lossless and lossy region) 
and $\phi=\frac{2\pi}{4m}+p\frac{\pi}{m}$
(cf.~Fig.~\ref{fig: scheeigen}),  respectively, 
%(center of lossy and lossless region), 
the power flow direction will change from CCW to CW 
and from CW to CCW, when tuning $\omega$ through the resonance, i.e., at resonance, there is no net CW or CCW power flow.
Similar behaviour can also be found for $\hat{\boldsymbol{\phi}}$ dipoles, where the power flow direction will change from CW (CCW) to CCW (CW) at $\phi=p\frac{\pi}{m}$ ($\phi=\frac{2\pi}{4m}+p\frac{\pi}{m}$, $p=0,1,2,...2m-1$).  
This gives one an external control method to change the directionality by simply changing the frequency.

In the current design, there is no gain materials involved. However, in general, as also mentioned in Ref.~\onlinecite{chen_revealing_2020}, one could also add gain material into the investigated ring resonator structure. 
%For example, if one replaces the lossy materials by gain materials, i.e., set $\delta n_{\rm Im}=-0.0012\sim-0.0002$ in Eq.~\eqref{eq: period_n}, 
%from full numerical dipole method, 
%we find 
%that one will found that it will have 
%exactly the opposite power flow compared with the ring with lossy materials.   
As an example, the results for the case with $\delta n_{\rm Im}=-0.001$ ($\delta n_{\rm Re}=0.001$) are shown in Fig.~\ref{fig: gain}. Two effective QNMs are found, whose total contribution to Purcell factors are net-negative (see Fig.~\ref{fig: gain} (a)), which are unphysical as recently demonstrated in Ref.~\onlinecite{franke_fermis_2021}, 
i.e., when the gain is present, the classical formula of the Purcell factor (Eq.~\eqref{eq: QNMpurcell}, here we refer it as $F_{\rm P}^{\rm class}$), and the traditional Fermi's golden rule for spontaneous emission decay do not work. However, the negative LDOS is correct and physical\footnote{Note a negative LDOS merely indicates that the net power flow around the LP dipole flows inwards, which can be the case even for a linear amplifier~\cite{PhysRevX.11.041020}}, and in agreement with the classical full numerical dipole result (red circles). 
To obtain the correct Purcell factor in amplifying dielectric media, a quantum mechanical fix is required as shown in Ref.~\onlinecite{franke_fermis_2021},
which results in (for a dipole with real dipole moment $\mathbf{d}$)
\begin{equation}\label{eq: PF_quant}
    F_{\rm P}^{\rm QM}({\bf r}_{\rm d})=F_{\rm P}^{\rm class}({\bf r}_{\rm d})+\frac{2\mathbf{d}\cdot\mathbf{K}^{\rm gain}(\mathbf{r}_{\rm d},\mathbf{r}_{\rm d},\omega)\cdot\mathbf{d}}{\hbar\epsilon_0\Gamma_0},
\end{equation}
where the second term is a positive gain-induced correction term, 
which will bring the total values to be correct net-positive $F_{\rm P}^{\rm QM}$.
In more detail, the general form of $\mathbf{K}^{\rm gain}(\mathbf{r}_{\rm d},\mathbf{r}_{\rm d},\omega)$ could be written as~\cite{franke_fermis_2021}
\begin{align}
\begin{split}
   \mathbf{K}^{\rm gain}(\mathbf{r}_{\rm d},\mathbf{r}_{\rm d},\omega)&\\
   =\int_{A_{\rm gain}}d\mathbf{s}&\big|{\rm Im}[\epsilon_{\rm gain}(\mathbf{s})]\big|\mathbf{G}(\mathbf{r}_{\rm d},\mathbf{s},\omega)\cdot\mathbf{G}^{*}(\mathbf{s},\mathbf{r}_{\rm d},\omega),
\end{split}
\end{align}
which are closely related with the Green function (Eq.~\eqref{eq: GFwithSUM_QNMs}), i.e., the two underlying QNMs, and the permittivity $\epsilon_{\rm gain}=n_{\rm gain}^2$ within the gain region $A_{\rm gain}$.
%\begin{align}
%    I^{\rm gain}_{\alpha}=&|{\rm Im}[\epsilon_{\rm gain}]|\int_{A_{\rm gain}}d\mathbf{r}|\tilde{f}_{\alpha,~y}(\mathbf{r})|^2,\\  
%     I^{\rm gain}_{12}=&|{\rm Im}[\epsilon_{\rm gain}]|\int_{A_{\rm gain}}d\mathbf{r}\tilde{f}_{1,~y}(\mathbf{r})\tilde{f}_{2,~y}^{*}(\mathbf{r}).
%\end{align}
Here we do not compute this quantum mechanical fix to the SE rate,
but further details are discussed in 
 Refs.~\onlinecite{franke_fermis_2021,ren_quasinormal_2021}.

In addition, the corresponding power flows from a linear dipole $\hat{\bf s}$ at $\phi=\pi/4m$ are shown in Fig.~\ref{fig: gain} (b-d).
Again, the QNM solution shows excellent agreement (both along CCW) with full numerical dipole solution, while the NM solution doesn't because the phases are ignored.
Moreover, compared with the ring with lossy materials (see Fig.~\ref{fig: QNMflow} (a), or Fig.~\ref{fig: compare}(a) and (e); along CW), the power flow with gain materials reverses to the opposite direction (Fig.~\ref{fig: gain} (b-c); along CCW), which is true for all the special cases such as those shown in Fig.~\ref{fig: QNMflow}, though here (Fig.~\ref{fig: gain}) we only show one of the examples. 
The reversed chiral power flow are closely related with the properties of the two underling QNMs with gain media, which are approximately differing $\pi$ on QNMs phases for $\tilde{\mathbf{f}}_{1/2}$, compared with those two QNMs with lossy media.
In addition, the imaginary parts of the corresponding QNM complex eigenfrequencies $\tilde{\omega}_{1/2}$ also change from negative to positive (gain mode). 
Taking these changes into account, the analytical power flow expression Eq.~\eqref{Eq:Spoyn} could exactly predict the reversed chirality. 
%\blue{SF: This reads great now!}

%\blue{SF: Shortly explain here why it is reversed.}
%\mygreen{JR: This is a great comment. Actually, it's not so obvious. I will think about it carefully.}
%\sh{Look at the analytical equations we gave and the change in QNM properties - this should explain it?}

%where we find exactly the opposite power flow from both QNM solution and full numerical dipole solution, compared with the ring with lossy materials

Also note that such chirality behavior is not found in regular ring resonators (without a refractive modulation) for LP $\hat{\bf s}$ and $\hat{\boldsymbol{\phi}}$ dipoles, i.e., no net power flow is obtained.
However, by using a circular-polarized dipole close to or inside the general ring resonator, chirality will exist, as shown in Ref.~\onlinecite{martin-cano_chiral_2019}, where positional dependent chirality for right- or left-handed circular dipoles are demonstrated. 
This is also similar to how one excites unidirectional 
propagation in photonic crystal waveguides~\cite{young_polarization_2015,Sllner2015}.
However, in all these cases, local symmetry breaking is possible by using
a circularly polarized dipole.\\

\section{Conclusions}\label{sec: conclusions}

We have investigated EP-like resonances
formed from index-modulated ring resonators,
where unusual chiral radiation from
LP dipoles
was recently observed~\cite{chen_revealing_2020}.
We showed that the full numerical dipole Purcell factor response
is quantitatively well explained in terms of the two dominating QNMs of this resonator, which could be negative in part of the frequency range, but the total contribution from two QNMs is always net-positive (for a lossy material system). In particular, 
when a LP $\hat{\bf s}$ dipole is located at $\phi=\pi/4m$, the analytical Poynting vectors obtained from the QNM propagator goes along the CW direction, which supports the experimental findings in Ref.~\onlinecite{chen_revealing_2020} for similar ring resonators.

Notably,
our explanation is in contrast with the view that the emitter does not couple to the eigenmodes of the system.
This is mainly because the QNMs are the correct eigenmodes and the perfect EP does not exists for our simulations, and the symmetry is naturally broken to allow chiral emission.
In addition, 
we also showed how such chirality is not limited to $\hat{\bf s}$ dipoles at $\phi=\frac{\pi}{4m}+p\frac{\pi}{m}$ ($p=0,1,2,...2m-1$), and we also demonstrated the opposite chirality for $\hat{\bf s}$ dipoles at $\phi=\frac{3\pi}{4m}+p\frac{\pi}{m}$ ($p=0,1,2,...2m-1$).  There is also similar chirality for linearly $\hat{\boldsymbol{\phi}}$ dipoles at these positions. All of these net power flows can be well explained and interpreted from the two underling QNMs and the corresponding Green function, where the QNM phases play a decisive and fundamental role on the light emission,
without having to invoke any unusual interpretation such as a missing dimension (Jordan vector).
In comparison, we also showed how a NM model fails to capture the correct power flow since it takes no account of the essential phases.
Moreover, frequency dependent chiral radiation was also investigated, which adds another degree of freedom to control the chirality in addition to the position and orientation of the emitter.

Lastly, we showed that when the lossy materials are replaced by gain materials in the index modulation, again two working classical QNMs are found, whose total contributions on the {\it classical Purcell factors} can be net-negative; this is a general failure of the classical theory,
but can be fixed to ensure a net-positive value through a quantum mechanical correction term~\cite{franke_fermis_2021}.  
In addition, the corresponding power flow direction will reverse compared to those with lossy materials.

\section*{Acknowledgements}
We  acknowledge funding from Queen's University,
the Canadian Foundation for Innovation, 
the Natural Sciences and Engineering Research Council of Canada, and CMC Microsystems for the provision of COMSOL Multiphysics.
We also acknowledge support from 
the Alexander von Humboldt Foundation through a Humboldt Research Award.
We thank Andreas Knorr
for discussions and  support.

\bibliography{refs}

\end{document}